\documentclass[pra,aps,twocolumn,superscriptaddress,a4paper,nofootinbib,showpacs]{revtex4}

\usepackage{graphicx,amssymb,amsmath}

\newcommand{\Tr}[0]{\ensuremath{\mathrm{Tr}}}

\begin{document}
\title{Observing a coherent superposition of an atom and a molecule}

\author{Mark~R.~Dowling} \email{dowling@physics.uq.edu.au}
\affiliation{School of Physical Sciences, The University of
  Queensland, Queensland 4072, Australia}%

\author{Stephen~D.~Bartlett}
\affiliation{School of Physics, The University of Sydney, New South
Wales 2006, Australia}%

\author{Terry~Rudolph}
\affiliation{Optics Section, Blackett Laboratory, Imperial College
London, London SW7 2BW, United Kingdom}%
\affiliation{Institute for Mathematical Sciences, Imperial College
  London, London SW7 2BW, United Kingdom}%

\author{Robert~W.~Spekkens}
\affiliation{Department of Applied Mathematics and Theoretical
Physics, University of Cambridge, Cambridge CB3
0WA, United Kingdom}%

\date{20 September 2006}

\begin{abstract}
We demonstrate that it is possible, in principle, to perform a
Ramsey-type interference experiment to exhibit a coherent
superposition of a single atom and a diatomic molecule. This
gedanken experiment, based on the techniques of Aharonov and
Susskind [Phys.~Rev.~\textbf{155}, 1428 (1967)], explicitly violates
the commonly-accepted superselection rule that forbids coherent
superpositions of eigenstates of differing atom number. A
Bose-Einstein condensate plays the role of a reference frame that
allows for coherent operations analogous to Ramsey pulses. We also
investigate an analogous gedanken experiment to exhibit a coherent
superposition of a single boson and a fermion, violating the
commonly-accepted superselection rule forbidding coherent
superpositions of states of differing particle statistics. In this
case, the reference frame is realized by a multi-mode state of many
fermions. This latter case reproduces all of the relevant features
of Ramsey interferometry, including Ramsey fringes over many
repetitions of the experiment. However, the apparent inability of
this proposed experiment to produce well-defined relative phases
between two distinct systems each described by a coherent
superposition of a boson and a fermion demonstrates that there are
additional, outstanding requirements to fully ``lift" the univalence
superselection rule.
\end{abstract}



\maketitle

\section{Introduction}

Part of the dogma of orthodox quantum mechanics is the presumed
existence of \emph{superselection rules}~\cite{WWW52,joos1996} for
certain quantities.  For instance, it is often stated that one
cannot create or observe quantum coherence between eigenstates of
differing charge, or of differing mass, or of number eigenstates of
particles with differing statistics (e.g., a superposition of a
boson and a fermion).\footnote{In Lorentz-invariant quantum field
theories, it has been argued that some superselection rules can be
derived within the theory~\cite{strocchi1974}. However, such
arguments do not apply to non-relativistic quantum theory, and in
particular do not apply when classical reference frames (i.e.,
measurement apparatuses) are employed within the theory.}
Originally, superselection rules were introduced to enforce
additional constraints to quantum theory beyond the well-studied
constraints of selection rules (conservation laws).  In a classic
paper, Aharonov and Susskind~\cite{aharonov1967} challenged the
necessity of such superselection rules, and outlined a gedanken
experiment for exhibiting a coherent superposition of charge
eigenstates as an example of how superselection rules can be
obviated in practice.

The gedanken experiment of Aharonov and Susskind, and subsequent
investigations~\cite{Mir69,shimizu2001,KMP04,BDSW06, terracunha2006}, highlighted
the requirement of an appropriate \emph{reference frame} in order to
exhibit coherence between eigenstates of superselected quantities.
For example, while a reference frame for spatial orientation is
required to exhibit coherence between states of differing angular
momentum (in some direction), a more exotic form of reference frame
is required to exhibit coherence between states of differing charge.
Arguably, it is the ubiquity of reference frames for some quantities
(such as spatial orientation or phase) and not for others (such
as the type of frame needed to exhibit superpositions of charge
eigenstates) that has led to the proposed superselection rules for
some quantities and not others.

While this gedanken experiment served to illustrate a concept,
recent advances in the preparation and manipulation of exotic
quantum states of matter may offer the opportunity to demonstrate
these concepts in experiment.  In this paper, we discuss the
principles of an experiment that may be performed with ultracold
atoms and molecules where the superselection rule in question is for
atom number, or equivalently, baryon number. (The existence of such
a superselection rule is commonly assumed,
cf.~\cite{WWW52,Cir96,Nar96,gardiner1997,Leg01,Ver03,Wis03}). The
experiment we present aims to exhibit quantum coherence between
states corresponding to a single atom and a diatomic molecule; the
reference frame in this case is a Bose-Einstein condensate (BEC).

We advance this concept further by outlining a gedanken experiment
to exhibit coherence between a boson and a fermion.  We demonstrate
that, using a reference frame consisting of many fermions in many
modes, the essential features in this experiment are similar to
those of the fully bosonic one, and that the commonly-accepted
superselection rule disallowing superpositions of a boson and a
fermion (the univalence superselection rule~\cite{Wig83}) can be
violated in principle.

Aharonov and Susskind used an operational approach to identify
coherence between eigenstates of some superselected quantity:  the
observation of Ramsey-type fringes in an interference experiment.
The reference frame allows for the implementation of Ramsey pulses
to create and subsequently measure the superposition states. We
follow this approach here, and extend their results by identifying
several salient features of the reference frame that allow for
high-visibility Ramsey fringes for many repetitions of the
experiment.  We demonstrate that such a reference frame can be
treated as either a classical or a quantum system, with both
descriptions leading to equivalent predictions. In addition, we
identify where in the mathematical formalism the coherent
superpositions arise: if the reference frame is treated as a quantum
system, this coherence arises in the relational degrees of freedom.

Finally, our gedanken experiment to exhibit a coherent superposition
of a boson and a fermion suggests that there must be an
\emph{additional} requirement beyond the ability to repeatedly
observe high-visibility interference fringes for one to say that a
superselection rule has been ``lifted.''  We show that this gedanken
experiment, despite possessing all of the salient features of a
Ramsey interferometry experiment, cannot induce a well-defined
\emph{relative} phase between two distinct systems each described by
a coherent superposition of a boson and a fermion. We therefore
demonstrate that the ability to repeatedly violate a superselection
rule is not equivalent to lifting it.

\section{Exhibiting quantum coherence with interferometry}

\label{sec:Ramsey}

Consider a two-level atom, defined by two energy eigenstates
$|g\rangle$ and $|e\rangle$.  How would one demonstrate coherence
between these two states?  That is, how does one discriminate the
coherent superposition $\frac{1}{\sqrt{2}}(|g\rangle + |e\rangle)$
from the incoherent mixture $\frac{1}{2}(|g\rangle\langle g| +
|e\rangle\langle e|)$?  A direct method would be to measure many
identically-prepared atoms in the basis $|\pm\rangle =
\frac{1}{\sqrt{2}}(|g\rangle \pm |e\rangle)$ and observe the
statistics.  However, in practice, preparations and measurements are
restricted to the basis $\{|g\rangle,|e\rangle\}$.  The standard
method, then, is to perform an interference experiment, in the form
of a Ramsey interferometer, which we now outline.

Between the preparation and the measurement, the interference
experiment makes use of two types of evolutions. The first evolution
is described by a Hamiltonian of the form
\begin{equation}
  \label{eq:PulseHam}
  \hat{H}_{\rm Ram} = \frac{\hbar\Omega}{2} \bigl(|g\rangle\langle e|
  + |e\rangle\langle g|\bigr) \,,
\end{equation}
where $\Omega$ is a real number.  This evolution is implemented by a
laser tuned to the energy difference of these two levels, and this
description is adequate if the laser pulse can be treated as an
external potential (i.e., with a well-defined amplitude and phase).
Application of this Hamiltonian for some finite time is called a
\emph{Ramsey pulse}, and if this Hamiltonian is applied for a time
$t=\pi/(2\Omega)$, the resulting unitary operation is called a
$\pi/2$-pulse.  The second type of evolution is governed by the free
Hamiltonian
\begin{equation}
  \label{eq:Phase}
  \hat{H}_{\rm free} = \Delta |e\rangle\langle e| \,,
\end{equation}
with $\Delta$ the detuning between the energy splitting between $|e\rangle$ and
$|g\rangle$ and the laser frequency when the laser is tuned off resonance.

\begin{figure}
\begin{centering}
\includegraphics[width=85mm]{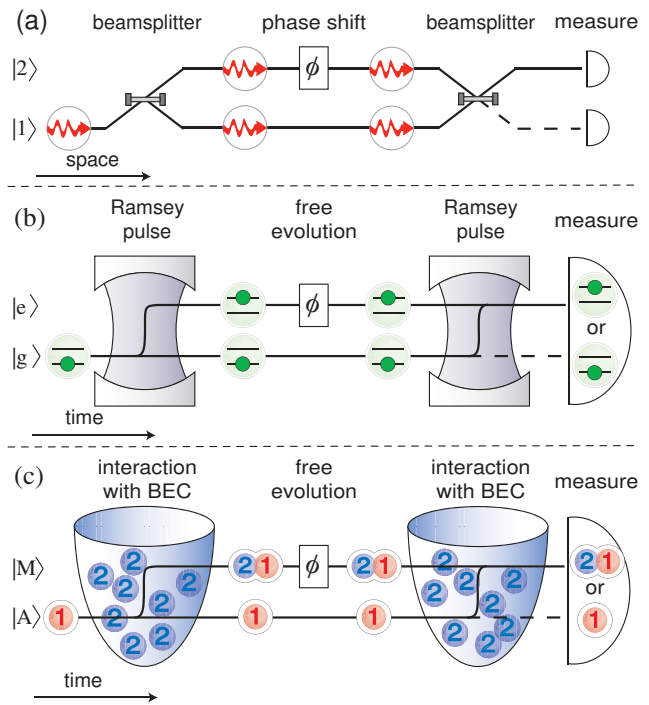}
\end{centering}
\caption{Schematic of the demonstration of quantum coherence in
different physical systems; (a) Mach-Zender interferometer, (b)
Ramsey interferometry, (c) the ultracold atom-molecule system
considered in this paper.} \label{fig:atommoleculefig}
\end{figure}

The interference experiment proceeds as follows.  The system is
prepared in the initial state $|g\rangle$, and then subjected to a
$\pi/2$-pulse.  The relative phase shift $\phi$ is then applied
using the second type of evolution for a time $\tau =
\phi/\Delta$, followed by another $\pi/2$-pulse.  At the time of
the preparation, and subsequent to each of the three interactions,
the state of the atom is, respectively,
\begin{align}
    \label{eq:R0}
    |\Psi_0\rangle &= |g\rangle \,, \\
    \label{eq:R1}
    |\Psi_1\rangle &= \tfrac{1}{\sqrt{2}}\bigl(|g\rangle
    - i |e\rangle \bigr) \,, \\
    \label{eq:R2}
    |\Psi_2\rangle &= \tfrac{1}{\sqrt{2}}\bigl(|g\rangle
    - i e^{-i \phi} |e\rangle \bigr) \,, \\
    \label{eq:R3}
    |\Psi_3\rangle &= \sin(\phi/2)|g\rangle
    -\cos(\phi/2)|e\rangle\,.
\end{align}
Finally, the atom is measured in the $\{|g\rangle, |e\rangle\}$
basis. By controlling the magnitude of the phase shift $\phi$
between the pulses over many runs of the experiment, one observes
so-called \emph{Ramsey fringes} -- oscillations in the probability
of measuring the outcome corresponding to $|g\rangle$ (or
$|e\rangle$) as a function of $\phi$.  Such Ramsey fringes are the
signature of coherent operation, i.e., that the description of the
system for the time period between the two $\pi/2$-pulses is given
by a coherent superposition of $|g\rangle$ and $|e\rangle$. If
instead the system was described at this intermediate time by an
incoherent superposition, then the resulting statistics would be
independent of $\phi$, i.e., no fringes would be observed.  See the
experimental paper of Bertet \textit{et al}.~\cite{haroche2001} for
a presentation of a Ramsey interferometry experiment that takes a
similar perspective to the one adopted in this paper.

The basic structure of this Ramsey interference experiment has
analogies in many other quantum systems.  An experiment using a
single photon and a Mach-Zender interferometer is formally
equivalent, with beamsplitters acting in the role of the Ramsey
pulses; see Fig.~\ref{fig:atommoleculefig}.  Other analogous
experiments are commonplace in atomic, molecular and nuclear
systems, and have more recently been demonstrated in artificial
structures such as semiconductor quantum dots~\cite{Bon98} and
superconducting qubits~\cite{Nak99}. Because the same abstract
structure can be realized in a wide variety of physical systems, the
identification between these different realizations of the same
basic interference experiment has been called the ``quantum Rosetta
stone''~\cite{Lee02}.  We seek to add another realization to this
list.

\section{Coherent superpositions of an atom and a molecule}
\label{sec:AtomMol}

We now consider an analog of the Ramsey experiment that aims to
exhibit coherence between a single atom and a molecule.  Consider a
bosonic atom; we denote the species of this atom as type 1.  Also,
consider a diatomic bosonic molecule, denoted $M$, which consists of
one atom of type 1 and one other bosonic atom of another type, 2. We
now define a two-level system, which will serve as the analogue in
our experiment of the two-level atom in the Ramsey experiment. This
two-level system is spanned by the following basis states:
\begin{equation}
    \label{eq:twolevelsystem}
  |A\rangle = |0\rangle_M |1\rangle_A  \,, \qquad
  |M\rangle = |1\rangle_M |0\rangle_A \,,
\end{equation}
where $|i\rangle_A$ is a Fock state for an atomic mode of type $1$
and $|i\rangle_M$ is a Fock state for a diatomic molecule mode.  One
may consider the single atom state, $|A\rangle$, as the analogue of
the ground state of the two-level atom in the Ramsey experiment, and
the molecule state $|M\rangle$ as the analogue of the excited state
of this two-level atom.  The aim is to demonstrate coherence in this
two-level system, i.e., coherence between $|A\rangle$ and
$|M\rangle$.  Such a demonstration of coherence violates a
superselection rule for atom number\footnote{We note that such an experiment is distinct from ones that aim to exhibit coherence between two atoms and a diatomic molecule, or between a BEC of N atoms and a BEC of N/2 diatomic molecules, as considered in~\cite{Hei00,Cal01,Don02}.}.

We will use a Bose-Einstein condensate (BEC) of atoms of type $2$ to
serve as our reference frame, i.e., as the analogue of the
electromagnetic field that constitutes the Ramsey pulse.  (The use
of a BEC as a phase reference has been discussed in~\cite{Dun99}.)
The analogue of the Ramsey pulse in our experiment is given by an
interaction of the two-level system with the reference frame (the
BEC). The relative phase shift will be implemented by free evolution
of the system. Using these two basic evolutions we will describe an
experiment that is formally equivalent to Ramsey interferometry.

\subsection{Using a quantum reference frame}
\label{subsec:QRF}

In the Ramsey interference experiment presented in
Sec.~\ref{sec:Ramsey}, the laser pulse which served as a reference
frame was treated as a external potential.  However, another
description of the same experiment could be presented wherein the
laser pulse was treated as a dynamical quantum system.  In general,
different descriptions of the same interference experiment are
possible depending on whether the reference frame is treated as a
non-dynamical macroscopic apparatus or as a dynamical quantum
system.  In the language of~\cite{BRS06} to treat the reference
frame as part of the apparatus is to use it as an \emph{external}
reference frame, while to treat it as a dynamical quantum object is
to use it as an \emph{internal} reference frame.

We begin by treating the reference frame -- in this case, the BEC --
as a fully-dynamical quantum system. At the end of Sec.~\ref{sec:relational}, we will
demonstrate that this experiment is equivalent to the Ramsey
interference experiment described in the previous section when the
reference frame (the BEC) is treated externally.

\subsubsection{Quantum state of the BEC}

For simplicity, we adopt a single-mode description of the BEC.  The
single mode corresponds to the Gross-Pitaevski ground state, which
has had great success in describing BEC
dynamics~\cite{Leg01,dalfovo1999}. Following the terminology
of~\cite{aharonov1967} we refer to the three modes -- the mode of
atom type 1, the molecular mode, and the BEC mode -- together as the
\emph{laboratory}\footnote{We will describe our proposed experiment
in terms of the dynamics of these three modes. There are, of course,
many details that are not captured by this simple model, for example
the many vibrational modes of the molecule, which would need to be
taken into consideration in an experiment.}. States of the
laboratory are most generally defined on the Fock space
$\mathcal{F}_M \otimes \mathcal{F}_A \otimes \mathcal{F}_2$ spanned
by the basis of Fock states $|n_M\rangle_M|n_A\rangle_A
|n_2\rangle_2$, where $n_M$, $n_A$ and $n_2$ are occupation numbers
for the modes.  In our experiment, $\hat{n}_M+\hat{n}_A$ will be a
constant of the motion.  Initially, it has eigenvalue $1$, as we
consider an initial state consisting of precisely one atom of type
$1$ and zero molecules. Thus, we can restrict our attention to the
two-dimensional subspace of $\mathcal{F}_M \otimes \mathcal{F}_A$
spanned by the two states $|A\rangle$ and $|M\rangle$ of
Eq.~\eqref{eq:twolevelsystem}.  We refer to this two-level system as
the \emph{system}, and the Hilbert space associated with it (spanned
by $|A\rangle$ and $|M\rangle$) is denoted $\mathcal{H}_S$. The
\emph{reference frame} is the remaining mode describing the BEC of atomic species
$2$, with infinite-dimensional Hilbert space $\mathcal{H}_{R} =
\mathcal{F}_2$.  States of the laboratory (system + reference frame)
are thus defined on $\mathcal{H}_S \otimes \mathcal{H}_{R}$. We will
use the modified Dirac notation $|\cdot \}$ for the state of the
reference frame, to emphasize the special role it plays.

The BEC consists of atoms of type $2$ in a single mode (the
Gross-Pitaevski ground state mentioned above).  In simplistic
treatments, it is common to treat the condensate as a coherent
state, i.e., to assign a state $|\beta\}$, defined in the number
basis as $|\beta\} = \sum_{n{=}0}^\infty c_n |n\}$ with $c_n =
\exp(-|\beta|^2/2)\beta^n/\sqrt{n!}$, as the quantum state of the
BEC.  However, as argued in~\cite{BRS06}, the coherence properties
of a state assigned to a bosonic mode are defined relative to a
classical reference frame for phase that is held in the background.
If the phase of the BEC is uncorrelated with any classical phase
reference in the background, as we will assume in our experiment,
then it is instead appropriate to assign it a quantum state that is
incoherently averaged over all possible orientations with respect to
the background phase reference. Such a state is given by
\begin{equation}\label{eq:coherentmixture2}
  \rho_0^{\rm rf} = \int_0^{2\pi}
  \frac{d\theta}{2\pi}e^{-i\hat{n}_2\theta}|\beta\}
  \{\beta| e^{i\hat{n}_2\theta} \,.
\end{equation}
It is straightforward to show that this state is equal to a
Poissonian mixture of number states
\begin{equation}\label{eq:coherentmixture}
  \rho_0^{\rm rf} = \sum_{n{=}0}^\infty p_n(\bar{n}) |n \}\{ n|\,,
\end{equation}
where $p_n(\bar{n}) = e^{-\bar{n}} \bar{n}^n/n!$ is a Poisson
distribution with $\bar{n}=|\beta|^2$.  It is this state,
$\rho_0^{\rm rf}$, that we choose to describe the BEC.\footnote{The
intuition that motivates using a coherent state for the BEC is the
belief that coherence is required to explain interference
experiments involving BECs such as~\cite{andrews1997}.  However,
coherent states are not required for such an explanation; it has
been demonstrated that number states, or incoherent mixtures of
number states such as Eq.~\eqref{eq:coherentmixture}, can also
interfere~\cite{javanainen1996}.  See~\cite{San03, Mol97, cable2005}
for further discussion of this fact.}  The treatment of the BEC as
an incoherent state is consistent with the arguments of
Refs.~\cite{javanainen1996,Cir96,Nar96}.

Eq.~\eqref{eq:coherentmixture2} is particularly useful as a
pedagogical tool for understanding interference experiments such as
the one described here:  the state $\rho_0^{\rm rf}$ can be
intuitively viewed as a coherent state $|\beta\}$, the phase of
which (arg$\beta$) is uncorrelated with any classical phase
reference used to describe the experiment.

If the BEC mode is initially described by this mixed state
$\rho_0^{\rm rf}$, and the system starts in the state $|A\rangle$, then the initial state of
the laboratory may be expressed as
\begin{equation}\label{eq:coherentmixturelabinitial}
  W_{0L} = |A\rangle\langle A| \otimes \rho_0^{\rm rf}
  = \int_0^{2\pi} \frac{d\theta}{2\pi}e^{-i\hat{N}_{\rm tot}\theta}
  |\Psi_0\rangle_L\langle\Psi_0| e^{i\hat{N}_{\rm tot}\theta} \,,
\end{equation}
where
\begin{equation}\label{eq:CSt0}
  |\Psi_0\rangle_L \equiv |A\rangle |\beta\}\,.
\end{equation}
and where the total atom number operator $\hat{N}_{\rm tot}$ is defined as
\begin{equation}\label{eq:TotalParticleNumber}
  \hat{N}_{\rm tot} \equiv 2\hat{n}_M+\hat{n}_A+\hat{n}_2\,.
\end{equation}
We define the \emph{twirling operator} acting on density operators
on the laboratory as
\begin{equation}
  \mathcal{T} [\rho_L] = \int_0^{2\pi} \frac{d\theta}{2\pi}e^{-i\hat{N}_{\rm tot}\theta}
  \rho_L e^{i\hat{N}_{\rm tot}\theta}\,.
\end{equation}
Then $W_{0L} = \mathcal{T}[|\Psi_0\rangle_L\langle\Psi_0|]$. It will
be illustrative to follow the evolution of the $\theta=0$ element
$|\Psi_0\rangle_L$ of the ensemble in
Eq.~\eqref{eq:coherentmixturelabinitial} through the experiment,
bearing in mind that the global phase $\theta$ is not physically
significant, and the quantum state of the laboratory is obtained by
averaging over this phase (implemented by the twirling operation
$\mathcal{T}$).

\subsubsection{Interactions with the quantum reference frame}

We now outline the Hamiltonians that will be used to induce the
required evolutions in our gedanken experiment.  The first is an
interaction between the system and the quantum reference frame. At a
Feshbach resonance, which occurs when an external magnetic field is
tuned so that the energy of two free atoms is equal to that of a
bound molecular state, coherent tunneling occurs between pairs of
atoms and molecules.  A simple model Hamiltonian for this phenomena,
where the two atoms are of different species
is~(cf.~\cite{corney2004})
\begin{equation}
  \label{eq:feshbachhamiltonian}
  \hat{H}_{\rm int} = \frac{\hbar\kappa}{2} (\hat{b}_M^\dagger \hat{b}_A \hat{b}_2 +
  \hat{b}_M \hat{b}_A^\dagger \hat{b}_2^\dagger )\,,
\end{equation}
where $\hat{b}_M$ is the annihilation operator for the bound
molecular state, and $\hat{b}_A$ and $\hat{b}_2$ are annihilation
operators for the modes containing the two distinct atomic
species.  This Hamiltonian can be
reexpressed in terms of an operator on the Hilbert space
$\mathcal{H}_S \otimes \mathcal{H}_{R}$ of the laboratory as
\begin{equation}
  \label{eq:feshbachhamiltonianPrime}
  \hat{H}_{\rm int}' = \frac{\hbar\kappa}{2}
  \bigl(|M\rangle\langle A| \otimes \hat{b}_2 +
  |A\rangle\langle M| \otimes \hat{b}_2^\dagger \bigr)\,.
\end{equation}

We will also make use of a Hamiltonian that induces a relative phase
shift on the system.  Such a Hamiltonian is provided by moving off
of the Feshbach resonance and allowing free evolution under the
Hamiltonian
\begin{equation}\label{eq:Hfree}
  \hat{H}_{\rm Free} = \hbar \omega_M \hat{n}_M +
  \hbar \omega_A \hat{n}_A + \hbar \omega_2 \hat{n}_2 \,,
\end{equation}
where $\hbar \omega_i$, $i = M,A,2$, are the internal energies of
the two atomic species and the molecule away from the Feshbach
resonance.  There will be some internal energy difference between
the bound molecular state and the sum of the two free atomic states
when off-resonance, given by
\begin{equation}
  \Delta_{\rm int} \equiv \omega_M - \omega_A - \omega_2\,.
\end{equation}
We note that both $\hat{N}_{\rm tot}$ and
$\hat{n}_M + \hat{n}_A$ are constants of the motion, and thus we can
move to an interaction picture in which the free-evolution
Hamiltonian on $\mathcal{H}_S$ is simply
\begin{equation}
  \label{eq:HFreePrime}
  \hat{H}'_{\rm Free} = \hbar\Delta_{\rm int}|M\rangle\langle M|\,,
\end{equation}
up to addition of a constant.

\subsubsection{The interference experiment}

We initiate the experiment with the laboratory in the state $W_{0L}$
of Eq.~\eqref{eq:coherentmixturelabinitial}. The system and
reference frame first interact for a time $t =
\pi/(2\kappa\sqrt{\bar{n}})$ at the Feshbach resonance according
to~\eqref{eq:feshbachhamiltonianPrime}. This interaction plays the
role of a Ramsey pulse.  As a result of this interaction, the state
of the laboratory evolves (in the Schr\"{o}dinger picture) to
$W_{1L} = \mathcal{T}[|\Psi_1\rangle_L\langle\Psi_1|]$, with
\begin{equation}
  \label{eq:CSt1}
  |\Psi_1\rangle_L = |A\rangle |\beta_A^1\} + |M\rangle |\beta_M^1\}
  \,,
\end{equation}
where we have defined unnormalized states
\begin{align}
  |\beta_A^1\} &\equiv \sum_{n=0}^\infty c_n
  \cos(\sqrt{\tfrac{n}{\bar{n}}}\tfrac{\pi}{4}) |n\}\,, \\
  |\beta_M^1\} &\equiv - i \sum_{n=0}^\infty c_n
  \sin(\sqrt{\tfrac{n}{\bar{n}}}\tfrac{\pi}{4}) |n{-}1\}\,.
\end{align}
We can interpret these states as (unnormalized) coherent states that
have undergone a disturbance due to the interaction with the system.

Next, we allow the laboratory to freely evolve for time $\tau$ away
from the Feshbach resonance according to~\eqref{eq:HFreePrime}. This
free evolution induces a relative phase between the atom and
molecule terms.  The result is a state $W_{2L} =
\mathcal{T}[|\Psi_2\rangle_L\langle\Psi_2|]$, with
\begin{equation}
  \label{eq:CSt2}
  |\Psi_2\rangle_L = |A\rangle | \beta_A^1\}
  + e^{-i \phi} |M\rangle |\beta_M^1\}\,,
\end{equation}
and where we have defined
\begin{equation}\label{eq:PhiDef}
    \phi \equiv \Delta_{\rm int} \tau \,,
\end{equation}
as the net relative phase shift between the atom and molecule terms
acquired during this stage.

Finally we implement the analogue of the second Ramsey pulse by
allowing the system and reference frame to interact again at the
Feshbach resonance for time $t = \pi/(2\kappa\sqrt{\bar{n}})$,
giving the final state $W_{3L} =
\mathcal{T}[|\Psi_3\rangle_L\langle\Psi_3|]$, with
\begin{equation}
  \label{eq:CSt3} |\Psi_3\rangle_L
  = |A\rangle |\beta_A^3\} + |M\rangle |\beta_M^3\}\,,
\end{equation}
where we have defined unnormalized states
\begin{align}
  |\beta_A^3\} &\equiv \sin(\phi/2)|\beta\}
  -i \cos(\phi/2) \sum_{n{=}0}^\infty
  c_n \cos(\sqrt{\tfrac{n}{\bar{n}}}\tfrac{\pi}{2}) \big)|n\}\,,  \\
  |\beta_M^3\} &\equiv -\cos(\phi/2)
  \sum_{n{=}0}^\infty c_n
  \sin(\sqrt{\tfrac{n}{\bar{n}}}\tfrac{\pi}{2} )|n{-}1\}\,,
\end{align}
and where we have ignored the global phase ($i e^{-i \Delta_{\rm
int} \tau/2}$) in the final states.  Finally, the system is measured
in the $\{|A\rangle,|M\rangle\}$ basis.

\subsubsection{A large-amplitude limit of the BEC}
\label{subsubsec:Limit}

We now examine the limit $\bar{n} \to \infty$.  The Poisson
distribution of the state~\eqref{eq:coherentmixture} has two key
properties that will be useful for our purposes.  First, in this
limit, the atom number distribution $p_n(\bar{n})$ of the BEC
becomes highly peaked about the mean atom number in the sense that
the standard deviation in atom number, $\Delta n = \sqrt{\bar{n}}$,
goes to zero relative to the mean.  This property ensures that the
interaction of the system with the BEC yields precisely a $\pi/2$
Ramsey pulse in this limit.\footnote{Each Fock state $|n\}$
component of the reference-frame state induces a Rabi oscillation
between $|A\rangle$ and $|M\rangle$ at frequency $\kappa\sqrt{n+1}$.
In the limit $\bar{n}\to\infty$, the fact that the uncertainty in
$n$ goes to zero relative to the mean ensures that all relevant
oscillations occur at the same frequency.} The second property is
that, in absolute terms, the standard deviation grows unbounded and
so intuitively we might think that the state of the BEC is
negligibly disturbed by the loss of a single atom in this limit.
Formally, the states of the reference frame after the first and
second interactions approach the following limits\footnote{These
limits follow from the fact that the inner products of the BEC
states with the appropriate coherent states approach unity in the
limit $\bar{n}\to\infty$.} as $\bar{n}\to\infty$
\begin{align}
  |\beta_A^1\} &\to \frac{1}{\sqrt{2}} |\beta\}\,, \
  &|\beta_A^3\} &\to \sin(\phi/2) |\beta\} \\
  |\beta_M^1\} &\to -\frac{i e^{i \arg\beta}}{\sqrt{2}} |\beta\}\,, \
  &|\beta_M^3\} &\to -e^{i \arg\beta} \cos(\phi/2)
  |\beta\}\,.
\end{align}
Therefore, in the limit $\bar{n}\to\infty$, the evolution of the
$\theta=0$ term of Eq.~\eqref{eq:coherentmixturelabinitial}, that
is, the evolution of $|\Psi_0\rangle_L$ given by
Eqns.~\eqref{eq:CSt0}, \eqref{eq:CSt1}, \eqref{eq:CSt2} and
\eqref{eq:CSt3}, reduces to
\begin{align}
    |\Psi_0\rangle_L &= |A\rangle |\beta\} \,, \\
    \label{eq:CSt1limit}
    |\Psi_1\rangle_L
    &\to \tfrac{1}{\sqrt{2}}\bigl(|A\rangle - i e^{i \arg\beta}|M\rangle\bigr)|\beta\} \,, \\
    \label{eq:CSt2limit}
    |\Psi_2\rangle_L
    &\to \tfrac{1}{\sqrt{2}}\bigl(|A\rangle - i e^{i(\arg\beta-\phi)}
    |M\rangle\bigr)|\beta\} \,, \\
    \label{eq:CSt3limit}
    |\Psi_3\rangle_L
    &\to \bigl(\sin(\phi/2)|A\rangle
    -e^{i \arg\beta} \cos(\phi/2)|M\rangle\bigr)|\beta\} \,.
\end{align}
We see that the $\theta=0$ term (in fact, any term) remains a
product on $\mathcal{H}_S \otimes \mathcal{H}_{R}$ for the entire
experiment. Moreover, applying the twirling operator $\mathcal{T}$
to such a product yields a separable state, which involves classical
correlations between the system and the reference frame but no
entanglement.  In addition, we note that in this limit the reduced
density operator of the reference frame remains $\rho_0^{\rm rf}$ of
Eq.~\eqref{eq:coherentmixture2} at each stage; i.e., the reference
frame is left undisturbed throughout the experiment.

The probabilities for detecting an atom or molecule at the end
display Ramsey oscillations
\begin{equation}\label{eq:perfectinterference}
  p_A = \sin^2(\phi/2)\,, \qquad p_M = \cos^2(\phi/2)\,.
\end{equation}
This interference pattern has perfect visibility.

\subsubsection{Interpreting the results}
\label{subsec:Interpreting}

We see therefore that the probabilities for detecting an atom or molecule exhibit
the familiar Ramsey oscillations, identified in
Sec.~\ref{sec:Ramsey} as the operational signature of coherence. So
it would seem that we can safely conclude that this experiment
demonstrates the possibility of a coherent superposition of an atom
and a molecule.

However, the careful reader might have noted the following peculiar
fact.  From Eqs.~\eqref{eq:CSt1limit} and \eqref{eq:CSt2limit}, we
find that the reduced density operators on $\mathcal{H}_S$ of
$|\Psi_2\rangle_{L}$ or $|\Psi_3\rangle_L$ are
\begin{align}
  &\tfrac{1}{\sqrt{2}}\bigl(|A\rangle - ie^{i\arg\beta}|M\rangle\bigr) \,, \\
  &\tfrac{1}{\sqrt{2}}\bigl(|A\rangle - ie^{i(\arg\beta-\phi)}|M\rangle\bigr)
  \,,
\end{align}
respectively. The reduced density operators on $\mathcal{H}_S$ of
$W_{2L}$ and $W_{3L}$ are obtained from these by averaging over
$\arg\beta$ through the use of the twirling operator $\mathcal{T}$.
In both cases, this state is found to be the completely mixed state
$\frac{1}{2}(|A\rangle\langle A| + |M\rangle\langle M|)$, i.e., an
\emph{incoherent} mixture of an atom and a molecule.  Thus, we have
a puzzle:  we have clearly predicted the standard operational
signature of coherence, namely Ramsey-type fringes, but the
coherence is not present in our mathematical description of the
system.

Before suggesting the resolution of this puzzle, we shall
demonstrate that it is not unique to the atom-molecule example we
are considering.  We find precisely the same peculiarity in the
context of the original Ramsey experiment, as follows.

Within the analysis of the Ramsey experiment presented in
Sec.~\ref{sec:Ramsey}, the electromagnetic field constituting the
Ramsey pulses was treated as an external potential.  As is well
known, this description is perfectly adequate if the fields have
large amplitude. But note that one could have also chosen to treat
this EM field \emph{within} the quantum formalism, as a dynamical
system interacting with the atom.  We emphasize that whether the
reference frame (the EM field) is treated internally or externally
is merely a choice of the physicist.  As long as the field has large
amplitude, either description is adequate to the task of making
accurate predictions about what will be observed by the
experimenter.

Treating the reference frame internally, we can take the state of
the field to be a coherent state of a single bosonic mode associated
with a bosonic annihilation operator $\hat{a}$.  The phase is
unimportant (i.e., the demonstration of Ramsey oscillations does not
depend on the phase of the coherent state), so one obtains the same
experimental predictions if this state is averaged over all
phases~\cite{Mol97,San03}, i.e., if one uses a quantum state for the
EM field of the same form as the BEC state of
Eq.~\eqref{eq:coherentmixture2}.

The relevant Hamiltonian for interacting the two-level atom with the
EM field in this internalized description is the Jaynes-Cummings
Hamiltonian
\begin{equation}
  \label{eq:JC}
  \hat{H}_{\rm JC} = \hbar\chi \bigl(|g\rangle\langle e| \otimes \hat{a}
  + |e\rangle\langle g| \otimes \hat{a}^\dag \bigr)  \,.
\end{equation}
Starting with the coherent state averaged over all phases for the EM
field and the Jaynes-Cummings interaction Hamiltonian, one is led to
a description of the original Ramsey experiment that is
\emph{precisely equivalent} to the one we have just provided for the
atom-molecule system.

Along with being formally equivalent to our atom-molecule
experiment, this description of the Ramsey experiment with an
internalized reference frame is operationally equivalent to the one
presented in Sec.~\ref{sec:Ramsey}, and necessarily makes the same
predictions. In particular, it agrees with the model of
Sec.~\ref{sec:Ramsey} in predicting the presence of Ramsey fringes.
Given that such a model is formally equivalent to the atom-molecule
experiment presented in the previous section, we see that that the
latter \emph{necessarily} predicted the presence of such fringes.

We now turn to the resolution of this puzzle, which applies to both
our proposed atom-molecule experiment as well as the Ramsey
experiment. We denote the system by $S$ and the reference frame by
$R$. In the case of the Ramsey experiment, $S$ is the atom and $R$
is the EM field constituting the Ramsey pulse, while in the case of
the atom-molecule interference experiment, $S$ is the mode pair of
atom mode and molecule mode and $R$ is the BEC constituting the
analogue of the Ramsey pulse.

In such an experiment, if the reference frame is treated externally,
the total Hilbert space is denoted $\mathcal{H}_S$ and the quantum
state on this Hilbert space describes \emph{the relation between $S$
and $R$}.  As demonstrated in Sec.~\ref{sec:Ramsey}, the observation
of fringes in this experiment implies coherence between states
$|g\rangle$ and $|e\rangle$ of this particular relational degree of
freedom.  Alternately, if the reference frame is treated internally,
the total Hilbert space is denoted by $\mathcal{H}_S \otimes
\mathcal{H}_R$; however, in this description the quantum state on
$\mathcal{H}_S$ describes the relation between $S$ and a
\emph{background} reference frame, distinct from $R$.  Thus although
it is standard practice to use a common notation, the Hilbert space
we denote by $\mathcal{H}_S$ when $R$ is treated externally and the
Hilbert space we denote by $\mathcal{H}_S$ when $R$ is treated
internally describe \emph{distinct} degrees of freedom.

Thus, it is a mistake to think that a coherent superposition of
states $|g\rangle$ and $|e\rangle$ on $\mathcal{H}_S$ when $R$ is
treated externally necessarily implies a coherent superposition of
states $|g\rangle$ and $|e\rangle$ on $\mathcal{H}_S$ when $R$ is
treated internally.  Specifically, if one investigates only the
reduced density matrix on $\mathcal{H}_S$ when $R$ is treated
internally, no coherence will be found because the reference frame
$R$ relative to which these coherences are defined has been
discarded in taking the partial trace.  To find the relation between
$S$ and $R$ when the latter is treated internally, one should not
look to the reduced density operator on $\mathcal{H}_S$ but rather
to the reduced density operator on a different Hilbert space: one
for which its degrees of freedom are the \emph{relation} between $S$
and $R$. In the following section, we identify this relational
Hilbert space and demonstrate that when Ramsey fringes are observed,
states in this Hilbert space are indeed coherent.

\subsection{A relational description}
\label{sec:relational}

We now provide the details of the relational description.  This
description could be viewed as an example of quantum coherence in
the presence of unobservable quantities, in this case the overall
phase, discussed in~\cite{nemoto2003}.  The averaging over all
phases, given by the twirling operator $\mathcal{T}$, ensures that
the density matrix of the laboratory is, at all times in this
experiment, block-diagonal in the eigenspaces of \emph{total} type-2
atom number $\hat{N}_2 = \hat{n}_2 + \hat{n}_M$.  (Note that this
total number operator also counts atoms of type 2 that are bound in
molecules.)  Thus, we can express the state of the laboratory at
each stage as
\begin{equation}\label{eq:labblockdiagonal}
  W_{iL} = \sum_{N{=}0}^\infty p_{N}(\bar{n})
  |\Psi^{(N)}_i\rangle_L\langle\Psi^{(N)}_i|\,, \ \ i=0,1,2,3,
\end{equation}
where
\begin{equation}
  |\Psi^{(N)}_i\rangle_L = \Pi_N |\Psi_i\rangle_L \,,
\end{equation}
with $\Pi_N$ the projector onto the eigenspace of $\hat{N}_2$ with
eigenvalue $N$, spanned by $|A\rangle|N\}$ and $|M\rangle|N-1\}$.

In the limit $\bar{n} \to\infty$, the states $W_{iL}$ have most of
their support on the subspaces for which $\bar{n} - \sqrt{\bar{n}}
\lesssim N \lesssim \bar{n} + \sqrt{\bar{n}}$, and using the same
approximations as in Sec.~\ref{subsubsec:Limit} we have, for $N$ in
this range,
\begin{align}
    |\Psi^{(N)}_0\rangle_L
    &\to |A\rangle |N\} \,, \\
    |\Psi^{(N)}_1\rangle_L
    &\to \tfrac{1}{\sqrt{2}}\bigl(|A\rangle |N\}
    - i |M\rangle|N{-}1\}\bigr) \,, \\
    |\Psi^{(N)}_2\rangle_L
    &\to \tfrac{1}{\sqrt{2}}\bigl(|A\rangle |N\}
    - i e^{-i \phi} |M\rangle |N{-}1\} \bigr) \,, \\
    |\Psi^{(N)}_3\rangle_L
    &\to \sin(\phi/2)|A\rangle |N\}
    -\cos(\phi/2)|M\rangle |N{-}1\}\,.
\end{align}
The fact that the coefficients in these superpositions are
independent of $N$ suggests that, in this limit, we can express
these states on an alternate Hilbert space, as we now demonstrate.

We define a new two-dimensional Hilbert space
$\mathcal{H}_{\text{rel}}$ with an orthonormal basis denoted by
$|A\rangle_{\text{rel}}$ and $|M\rangle_{\text{rel}}$, corresponding
respectively to $n_M=0$ (no molecules) and $n_M=1$ (one molecule).
We call this the \emph{relational} Hilbert space. We also define a
new Hilbert space $\mathcal{H}_{\text{gl}}$ which has an orthonormal
basis labeled by $N$ and defined for $N \ge 1$. We call this the
\emph{global} Hilbert space.  Define the subspace $\mathcal{H}'_L$
of $\mathcal{H}_L$ as the orthogonal complement to the vector
$|A\rangle|0\}$.  We can define a linear map from the subspace
$\mathcal{H}'_L$ of $\mathcal{H}_S \otimes \mathcal{H}_R$ to this
new tensor product Hilbert space $\mathcal{H}_{\rm rel} \otimes
\mathcal{H}_{\rm gl}$ by its action on basis vectors as
\begin{align} \label{eq:tensormap}
  |A\rangle |N\} &\mapsto |A \rangle_{\rm rel} |N\rangle_{\rm gl} \,, \\
  \label{eq:tensormap2}
  |M\rangle |N-1\} &\mapsto |M \rangle_{\rm rel} |N\rangle_{\rm gl} \,,
\end{align}
for all $N\ge1$.

It is illustrative to construct this alternate Hilbert space and the
associated map~(\ref{eq:tensormap}-\ref{eq:tensormap2}) by
simultaneously diagonalizing two commuting operators. Note that the
states $|A\rangle|N\}$ and $|M\rangle|N\}$ on $\mathcal{H}_S\otimes
\mathcal{H}_R$ are simultaneous eigenstates of the operators
$\hat{n}_M$ (or equivalently, $\hat{n}_1$) and $\hat{n}_2$; the
former labels states on $\mathcal{H}_S$, and the later labels states
on $\mathcal{H}_R$. Specifically,
\begin{align}
  |A\rangle|N\}&=|n_M{=}0,n_2{=}N\rangle \\
  |M\rangle|N\}&=|n_M{=}1,n_2{=}N\rangle\,.
\end{align}

We can instead choose a different set of commuting operators to
achieve an alternate tensor product structure for the laboratory
Hilbert space.  We choose the commuting operators $\hat{n}_M$ and
$\hat{N}_2 = \hat{n}_2 + \hat{n}_M$, the latter being the total
number operator for atoms of type 2. We note that the states
$|A\rangle|N\}$ and $|M\rangle|N\}$ are also joint eigenstates of
$\hat{n}_M$ and $\hat{N}_2$, so that we may write
\begin{align}
  |A\rangle|N\}&=|n_M=0,N_2=N\rangle \\
  |M\rangle|N\}&=|n_M=1,N_2=N+1\rangle\,.
\end{align}

We note that in the limit $\bar{n}\rightarrow \infty$, the states we
consider have no support on the vector
$|A\rangle|0\}=|n_M{=}0,N_2{=}0\rangle$ and thus we can focus our
attention on the subspace $\mathcal{H}'_L$ of $\mathcal{H}_L$ that
is orthogonal to this vector.  The states on $\mathcal{H}'_L$ are of
the form $|n_M,N_2\rangle$ with $N_2 \ge 1$.

Because the spectra of $\hat{n}_M$ and $\hat{N}_2$ are independent,
we can introduce a new tensor product structure
$\mathcal{H}_{\text{rel}}\otimes\mathcal{H}_{\text{gl}}$ on
$\mathcal{H}'_L$ which is made by identifying
\begin{align}
  |A\rangle_{\text{rel}}|N\rangle_{\text{gl}}
  &\equiv |n_M{=}0,N_2{=}N\rangle\,, \\
  |M\rangle_{\text{rel}}|N\rangle_{\text{gl}}
  &\equiv |n_M{=}1,N_2{=}N\rangle\,,
\end{align}
for all $N\ge 1$.  We then have a vector space isomorphism
\begin{equation}
  \mathcal{H}'_L \cong
  \mathcal{H}_{\text{rel}}\otimes\mathcal{H}_{\text{gl}}\,,
\end{equation}
This identification recovers the map of
Eq.~(\ref{eq:tensormap}-\ref{eq:tensormap2}).

Note that, under the map of
Eq.~(\ref{eq:tensormap}-\ref{eq:tensormap2}), we have
\begin{equation}\label{eq:MapEntanglementToCoherence}
    \alpha|A\rangle|N\} + \beta|M\rangle|N{-}1\} \mapsto
    \bigl( \alpha|A\rangle_{\rm
    rel} + \beta|M\rangle_{\rm rel}\bigr) |N\rangle_{\rm gl} \,,
\end{equation}
so that while the reduced density operator for this state on the
system $\mathcal{H}_S$ is an \emph{incoherent} mixture of
$|A\rangle$ and $|M\rangle$, the reduced density operator on the new
subsystem $\mathcal{H}_{\text{rel}}$ is a \emph{coherent}
superposition of $|A\rangle_{\rm rel}$ and $|M\rangle_{\rm rel}$.

This fact implies that the states $W_{iL}$ of the laboratory (in the
limit $\bar{n}\to\infty$) map to product states
\begin{equation}
  W_{iL} = |\Psi_i\rangle_{\rm rel} \langle \Psi_i| \otimes
  \rho_{\rm gl} \,,
\end{equation}
where
\begin{align}
    \label{eq:t0rel}
    |\Psi_0\rangle_{\rm rel}
    &= |A\rangle_{\rm rel} \,, \\
    \label{eq:t1rel}
    |\Psi_1\rangle_{\rm rel}
    &= \tfrac{1}{\sqrt{2}}\bigl(|A\rangle_{\rm rel}
    - i |M\rangle_{\rm rel} \bigr) \,, \\
    \label{eq:t2rel}
    |\Psi_2\rangle_{\rm rel} &= \tfrac{1}{\sqrt{2}}\bigl(|A\rangle_{\rm rel}
    - i e^{-i \phi} |M\rangle_{\rm rel} \bigr) \,, \\
    \label{eq:t3rel}
    |\Psi_3\rangle_{\rm rel}  &= \sin(\phi/2)|A\rangle_{\rm rel}
    -\cos(\phi/2)|M\rangle_{\rm rel}\,,
\end{align}
and
\begin{equation}\label{eq:BosonRhoGL}
    \rho_{\rm gl}
    = \sum_{N=1}^\infty p_{N}(\bar{n}) |N\rangle_{\rm gl}\langle N|\,.
\end{equation}

This new tensor product structure demonstrates explicitly how we
resolve the puzzle posed in the previous section. Both the states of
$\mathcal{H}_S$ and the states of $\mathcal{H}_{\text{rel}}$ are
labeled by the number of molecules, and consequently describe
whether the system is an atom or a molecule. Thus, the question
``can one have a coherent superposition of an atom and a molecule?''
is seen to be ambiguous as stated. Does it refer to a coherent
superposition of $|A\rangle$ and $|M\rangle$ on $\mathcal{H}_S$ or
to a coherent superposition of $|A\rangle_{\text{rel}}$ and
$|M\rangle_{\text{rel}}$ on $\mathcal{H}_{\text{rel}}$?  To resolve
the ambiguity, we take an operational stance.  By arguing in analogy
with the Ramsey interference experiment, we have proposed that an
operational signature of coherence is the appearance of Ramsey
fringes, and we have shown that this coincides with having a
coherent superposition of $|A\rangle_{\text{rel}}$ and
$|M\rangle_{\text{rel}}$ on $\mathcal{H}_{\text{rel}}$.

We finish our analysis by considering the dynamics on the laboratory
in terms of the new tensor product structure. Using the map of
Eq.~(\ref{eq:tensormap}), we find that the free Hamiltonian for the
evolution between Ramsey pulses, defined by
Eq.~\eqref{eq:HFreePrime}, becomes simply
\begin{equation}\label{eq:HFreePrimeRel}
  \hat{H}'_{\mathrm{Free}} = \hbar\Delta_{\rm int}\left\vert
  M\right\rangle _{\mathrm{rel}}\left\langle M\right\vert\,,
\end{equation}
and the Hamiltonian governing the interaction between the system and
the reference frame, Eq.~\eqref{eq:feshbachhamiltonianPrime}, becomes
\begin{equation}
  \hat{H}'_{\mathrm{int}}=\frac{\hbar\kappa}{2}\bigl(
  |M\rangle_{\mathrm{rel}}\langle A|+|A\rangle_{\mathrm{rel}}\langle
  M|\bigr)\otimes \sqrt{\hat{N}_{\mathrm{gl}}}\,.
\end{equation}

The effective map for the interaction with the BEC is determined as
follows. Noting that the interaction occurs for a time
$t=\pi/2\kappa\sqrt{\bar{n}}$ and that the initial state on
$\mathcal{H}_{\mathrm{gl}}$ is $\rho_{\mathrm{gl}}$ of
Eq.~\eqref{eq:BosonRhoGL}, the effective evolution on
$\mathcal{H}_{\rm rel}$ is represented by a completely-positive
trace-preserving map $\mathcal{E}$ of the form
\begin{align}
  \mathcal{E}(\rho_{\mathrm{rel}})
  &=\mathrm{Tr}_{\mathrm{gl}}\bigl(
  U_{\mathrm{int}}(\rho_{\mathrm{rel}}\otimes
  \rho_{\mathrm{gl}})U_{\mathrm{int}}^{\dag}\bigr) \nonumber  \\
  &=\sum_{N}p_{N}(\bar{n}){}_{\rm gl}\langle N| U_{\mathrm{int}}
  \rho_{\mathrm{rel}}U_{\mathrm{int}}^{\dag}|N\rangle_{\mathrm{gl}}
  \nonumber \\
  & =\sum_{N}p_{N}(\bar{n})e^{i \hat{H}^{(N)}_{\rm int}t}
  \rho_{\mathrm{rel}}e^{-i\hat{H}^{(N)}_{\rm int}t} \,.
\end{align}
where $\hat{H}^{(N)}_{\rm int}={}_{\rm gl}\langle
  N|\hat{H}'_{\mathrm{int}}|N\rangle_{\mathrm{gl}}$. But given that in the limit of
large $\bar{n},$ the distribution $p_{N} (\bar{n})$ is only
significant in the range $\bar{n}-\sqrt{\bar{n}}\lesssim
N\lesssim\bar{n}+\sqrt{\bar{n}}$ and given that $\lim_{\bar{n}\to\infty
}\frac{(\bar{n}\pm\sqrt{\bar{n}})t}{\bar{n}t}=1,$ we have that in this limit
\begin{equation}
  \mathcal{E}(\rho_{\mathrm{rel}})=e^{i\hat{H}_{\mathrm{Ram}}t}
  \rho_{\rm rel}e^{-i\hat{H}_{\mathrm{Ram}}t} \,,
\end{equation}
where
\begin{equation}\label{eq:BosonRelRamsey}
  \hat{H}_{\mathrm{Ram}}=\frac{\hbar\kappa\sqrt{\bar{n}}}{2}\bigl(
  |M\rangle_{\mathrm{rel}}\langle A| +|A\rangle_{\mathrm{rel}}\langle
  M|\bigr)\,.
\end{equation}
This has a natural interpretation as the analogue of a Ramsey pulse
where the pulse is implemented by a BEC that is treated as a
classical external field.

We can now make a complete comparison of our relational description
of the atom-molecule interference experiment with the original
description of the Ramsey experiment in Sec.~\ref{sec:Ramsey} (where
the fields corresponding to the Ramsey pulses were treated as
external potentials). The Hamiltonians
$\hat{H}_{\mathrm{Free}}^{\prime}$ of Eq.~\eqref{eq:HFreePrimeRel}
and $\hat{H}_{\mathrm{Ram}}$ of Eq.~\eqref{eq:BosonRelRamsey}
governing the relational degree of freedom are precisely analogous
to the Hamiltonians of Eq.~\eqref{eq:Phase} and
Eq.~\eqref{eq:PulseHam} governing the internal state of the atom in
the Ramsey experiment, and the states on
$\mathcal{H}_{\mathrm{rel}}$ at the four stages of the experiment,
given by Eqs.~(\ref{eq:t0rel})-(\ref{eq:t3rel}), are precisely
analogous to the states for Eqs.~(\ref{eq:R0})-(\ref{eq:R3}) for the
internal states of the atom in the Ramsey experiment. We conclude
that $\mathcal{H}_{\mathrm{rel}}$ can be understood as describing
either: (i) the relation between the system (atom and molecule
modes) and the reference frame formed by the BEC when the latter is
treated internally, or (ii) the system (atom and molecule modes)
when the reference frame formed by the BEC is treated as an external
potential.

\section{Coherent superpositions of a boson and a fermion}
\label{sec:bosonfermioncoherence}

We now repeat this analysis for the case when the atoms are
fermions, and demonstrate that it is possible in principle to
exhibit coherence between a fermion (a single atom) and a boson (a
molecule).  This result is quite surprising; it is commonly accepted
that there exists a superselection rule preventing a coherent
superposition of a boson and a fermion. We emphasize that we are
considering a superposition of a single fermionic atom with a single
bosonic molecule, not a superposition of two fermionic atoms and a
composite bosonic molecule as considered in~\cite{Reg03,Jav04}.

Consider our laboratory to consist of two types of atomic species,
type 1 and type 2, which are fermions, along with a bosonic diatomic
molecule consisting of one of each type of fermion. As with the
previous discussion, we will consider creating a superposition of an
atom of type 1 and a molecule, using atoms of type 2 as a reference
frame.

As we are using fermions, the natural Hilbert space for states of
the laboratory will be a Fock space.  However, we will want to make
use of a \emph{tensor product structure} of the laboratory Hilbert
space which divides it into a system and a reference frame (and,
subsequently, into a relational and a global Hilbert space). To do
this, we will make use of the natural mapping between the Fock space
$\mathcal{F}^N$ of $N$ fermionic modes and the tensor-product
Hilbert space $(\mathbb{C}^2)^{\otimes N}$ of $N$ qubits, given by
Bravyi and Kitaev~\cite{Bra02} in the Fock basis as
\begin{equation}\label{eq:LFMtoQubits}
    |n_1,n_2,\ldots,n_N\rangle \mapsto |n_1\rangle \otimes
    |n_2\rangle \otimes \cdots \otimes |n_N\rangle \,,
\end{equation}
for $n_i \in \{0,1\}$.  This identification implies a non-trivial
relation between operations on $\mathcal{F}^N$ and operations on
$(\mathbb{C}^2)^{\otimes N}$, as a result of phases acquired by
commuting operations through occupied modes.  Fortunately, due to
the highly-incoherent nature of the states that we will make use of,
and by working with mixed rather than pure states, we will find that
this non-trivial identification does not add much additional
complication.

Again, we use the notation
\begin{equation}
  |A\rangle = |0\rangle_M |1\rangle_A \,, \qquad
  |M\rangle = |1\rangle_M |0\rangle_A \,,
\end{equation}
where the first mode $|\cdot\rangle_M$ is bosonic (the molecule), and the second mode $|\cdot\rangle_A$ is fermionic (atom type 1). We define a \emph{system} Hilbert space $\mathcal{H}_S$ spanned by these two
states $|A\rangle$ and $|M\rangle$.

\subsection{Using a quantum reference frame consisting of fermions}
\label{subsec:FermionQRF}

\subsubsection{Generalizing the previous experiment}

In order to construct an interference experiment that exhibits
coherence between a boson and a fermion, we must identify an
appropriate reference frame consisting of fermions.  This is
non-trivial, given the difficulties of defining a fermionic coherent
state that has analogous properties to the standard (bosonic)
coherent state~\cite{Cah99,Gri02,Tyc05}. Again, we meet these
challenges by being operational.

We first examine some of the properties of the state of the BEC in
the experiment discussed in Sec.~\ref{subsec:QRF} which allowed it
to serve as a good reference frame.  To obtain good visibility in
the experiment, we required that all relevant Rabi oscillations
corresponding to different Fock state components $|n\rangle$ in the
state of the reference frame occur at the same frequency, which is
obtained by requiring the variance in $n$ to be small compared to
the mean, $\bar{n}$.  This requirement could be satisfied by a state
with a modest value of $\bar{n}$, and we could still have predicted
good fringe visibility. For example, a single Fock state $|n\rangle$
with $n\geq 1$ satisfies this requirement.

In addition, though, we noted that the reduced density operators for
the reference frame at each state of the experiment were undisturbed
in the limit $\bar{n} \rightarrow \infty$, which would allow for the
experiment to be repeated many times using the same reference frame.
For this additional condition to be satisfied, a large absolute
variance in total atom number is required (thus implying a large
mean total atom number).

We now consider an analogous situation in the fermion case.  First,
consider a reference frame of fermionic atoms that consists of a
single mode.  Due to the Pauli exclusion principle, the mean atom
number of this reference frame can be at most one.  Consider using a
Hamiltonian of the form
\begin{equation}
  \label{eq:Hfermion}
  \hat{H} = \frac{\hbar\kappa}{2} \left( \hat{b}_M^\dagger \hat{f}_{2}
  \hat{f}_A + \hat{f}_A^\dagger \hat{f}_{2}^{\dagger} \hat{b}_M \right)\,,
\end{equation}
where $\hat{b}_M$ is the boson annihilation operator for the
molecular mode, $\hat{f}_A$ is the fermion annihilation operator for
the mode of atom type $1$, and $\hat{f}_2$ is the annihilation
operator for the reference-frame mode of atom type 2.  It is clear
that if the initial state of the system is $|A\rangle$ and of the
reference frame is $|1\}$, then it is possible to perform an
interference experiment yielding maximum visibility.

However, in this experiment, the state of the reference frame is
highly disturbed.  One might (rightly) argue that it is essentially
just a Rabi oscillation.  We now consider what conditions on the
state of the reference frame must be satisfied in order for it to be
undisturbed throughout the experiment, thereby making it analogous
to the bosonic case in this regard as well.  First, we require that
the mean number of fermions in the reference frame must be large. To
achieve this, because of the Pauli exclusion principle, the
reference frame must be multi-mode.  Second, we cannot allow every
new system to interact with the same mode of the reference frame.
The reason for the latter is that a single mode can have at most one
fermion, and if we happen to find a molecule at the end of the
Ramsey experiment (and we should assume the worst and say that we do
find a molecule) then this fermion has been depleted from that mode.
The next system that interacts with this mode will therefore only
interact with the vacuum.

We now define a multi-mode reference frame with a large mean number
of atoms, and an associated multi-mode interaction between the
system and the reference frame.  The experiment we describe yields
high visibility and also leaves the state of the reference frame
undisturbed.

\subsubsection{The quantum state of the reference frame}
\label{subsec:FermionQRFstate}

Consider an $K$-mode fermionic reference frame of atoms of type 2,
using the tensor product structure of Eq.~\eqref{eq:LFMtoQubits},
initially prepared in the state
\begin{equation}
  \label{eq:multimodefermioninitial}
  \rho_0^{\rm rf} = \sigma^{\otimes K}\,,
\end{equation}
where $\sigma$ is given by
\begin{equation}
  \label{eq:singlefermionreference frame}
  \sigma = \epsilon |0\}\{0|+(1-\epsilon)|1\}\{1|\,.
\end{equation}
The distribution of total atom number in the state $\rho_0^{\rm rf}$
is given by the binomial distribution $c^K_n(1{-}\epsilon)$, where
we define the binomial coefficient $c^K_n(p)$ of $n$ successes in
$K$ trials where the probability of success is $p$ as
\begin{equation}
  \label{eq:binomialdef} c^K_n(p) = \tbinom{K}{n} p^n(1-p)^{K-n}\,.
\end{equation}
As we will argue below, in the limit $K\to\infty$ this atom number
distribution has the property that the loss of a single particle
leaves the distribution indistinguishable from the original.

\subsubsection{The interference experiment}

We initiate the experiment with the system prepared in the state
$|A\rangle$ and the reference frame in the state $\rho_0^{\rm rf}$
of Eq.~\eqref{eq:multimodefermioninitial}.  The initial state of the
laboratory is thus
\begin{equation}\label{eq:FermionLabInitial}
    W_{0L} = |A\rangle\langle A| \otimes \rho_0^{\rm rf} \,.
\end{equation}
To perform the operation that is analogous to a Ramsey $\pi/2$-pulse
in this experiment, we use the Hamiltonian
\begin{equation}
  \label{eq:Hfermions}
  \hat{H}^{(j)} = \frac{\hbar\kappa}{2} \left( \hat{b}_M^\dagger \hat{f}_{2}^{(j)}
  \hat{f}_A + \hat{f}_A^\dagger \hat{f}_{2}^{(j)\dagger} \hat{b}_M \right)\,,
\end{equation}
where $\hat{b}_M$ is the boson annihilation operator for the
molecular mode, $\hat{f}_A$ is the fermion annihilation operator for
the mode of atom type $1$, and $\hat{f}^{(j)}_2$ is the annihilation
operator for the $j$th mode ($j=1,2,\ldots,K$) of the reference
frame for atom type 2. Again, this Hamiltonian describes the
evolution at a Feshbach resonance, at which the energy of two free
fermions (one each of type 1 and 2) is equal to the energy of a
bound molecular state.

One might naturally consider implementing the $\pi/2$-pulses by
using the unitary operation generated by
\begin{equation}\label{eq:CoherentFermionFeshbach}
    \hat{H}_{\rm coh} = \sum_{j} \hat{H}^{(j)}\,.
\end{equation}
The problem with such a Hamiltonian is that it only couples the
system with a \emph{single} reference-frame mode, specifically, the
mode associated with the fermionic annihilation operator $\hat{F} =
\tfrac{1}{\sqrt{M}} \sum_j \hat{f}_k$.  As discussed above, such a coupling
is unsatisfactory.

Consider instead the following evolution: the system interacts with
a \emph{random} mode $j$ of the reference frame via the interaction
$\hat{H}^{(j)}$.  We can formalize such an evolution by using the
language of quantum operations~\cite{nielsen2000a}.  Let
$U_{\pi/2}^{(j)}$ be the unitary operation that describes
$\pi/2$-pulse obtained by coupling the system and the $j^{\rm th}$
reference-frame mode via the Hamiltonion $\hat{H}^{(j)}$ for a time
$t = \pi/(2\kappa)$.\footnote{Note that the interaction times do not
scale inversely with the square root of the mean atom number as only
one reference-frame mode containing at most one fermion interacts.
Essentially, the system undergoes a type of Rabi oscillation with a
single mode of the reference frame.}  Let $U_\phi$ be the unitary
operation that applies a phase shift $\phi= \Delta_{\rm int}\tau$ to
the $|M\rangle$ component of the system, i.e., evolution for time
$\tau$ according to the Hamiltonian $\hat{H}_{\rm int} = \Delta_{\rm
int}|M\rangle\langle M|$.

If laboratory is initially in the state $W_{0L}$, and if the system
interacts with a \emph{known} reference-frame mode $j$, the state at
each stage of the experiment is given by
\begin{align}
  \label{eq:fermionU1}
  W^{(j)}_{1L} &=  U_{\pi/2}^{(j)} W_{0L} (U_{\pi/2}^{(j)})^{-1} \,, \\
  \label{eq:fermionU2}
  W^{(j)}_{2L} &=  U_\phi U_{\pi/2}^{(j)} W_{0L} (U_\phi U_{\pi/2}^{(j)})^{-1} \,, \\
  \label{eq:fermionU3}
  W^{(j)}_{3L} &=  U_{\pi/2}^{(j)} U_\phi U_{\pi/2}^{(j)} W_{0L}
  (U_{\pi/2}^{(j)} U_\phi U_{\pi/2}^{(j)})^{-1}\,,
\end{align}
where the unitaries $U^{(j)}$ are taken to act as the identity on
all modes other than the $j$th mode.

To describe an evolution where, with probability $1/K$, the system
interacts with the $j^{\rm th}$ reference-frame mode, we make use of
generalized quantum operations, which have the form of
completely-positive trace-preserving (CPTP) maps given, in this
instance, by convex combinations of unitary transformations.  We
describe the evolution that takes the laboratory from the initial
state $W_{0L}$ to a state $W_{iL}$ for steps $i=1,2,3$ as
\begin{align}
  \label{eq:fermionW1}
  W_{1L} &= \mathcal{E}_1(W_{0L}) =
  \frac{1}{K} \sum_{j=1}^K  U_{\pi/2}^{(j)} W_{0L} (U_{\pi/2}^{(j)})^{-1} \,, \\
  W_{2L} &= \mathcal{E}_2(W_{0L}) \nonumber \\
  &\qquad =
  \frac{1}{K} \sum_{j=1}^K U_\phi U_{\pi/2}^{(j)} W_{0L} (U_\phi U_{\pi/2}^{(j)})^{-1} \,,
  \label{eq:fermionW2}
  \\
  W_{3L} &= \mathcal{E}_3(W_{0L}) \nonumber \\
  &\qquad =
  \frac{1}{K} \sum_{j=1}^K U_{\pi/2}^{(j)} U_\phi U_{\pi/2}^{(j)} W_{0L}
  (U_{\pi/2}^{(j)} U_\phi U_{\pi/2}^{(j)})^{-1}\,.
  \label{eq:fermionW3}
\end{align}
As noted above, there is a non-trivial relation between a unitary
$U_{\pi/2}^{(j)}$ coupling two modes together in the tensor product
Hilbert space, and the same coupling on the Fock space, due to the
phases acquired by commuting through modes $j'$ for $j'<j$; in
general, this mapping can be determined using the techniques
of~\cite{Bra02}.  However, this mapping does not exhibit any
non-trivial consequences for the interaction presented here, for the
following simple reason.  Note that each sequence of unitary
operations (either $U_{\pi/2}^{(j)}$, $U_\phi U_{\pi/2}^{(j)}$ or
$U_{\pi/2}^{(j)} U_\phi U_{\pi/2}^{(j)}$) only couples the system
with a single reference-frame mode $j$; such coupling will thus lead
to a non-trivial phase due to the modes $j'<j$. However, the
expressions above sum \emph{incoherently} over the different
possibilities $j$, and thus the phases acquired for each term in
this sum do not interfere.

For clarity, it will be useful to follow the evolution associated
with the $j=1$ element of the above operation, remembering at each
step that the state of the laboratory is described by interacting
the system with a random mode.  To do this, we define the
\emph{shuffling operation} $\mathcal{S}$, which is the incoherent
symmetrizer acting on states of the reference frame as
\begin{equation}
  \label{eq:Shuffling}
  \mathcal{S}\left[\rho^{\rm rf} \right]
  = \frac{1}{K!} \sum_{\pi} S_\pi \rho^{\rm rf} S_\pi^\dagger \,,
\end{equation}
where the sum is over all permutations $\pi$ of $K$ indices and
$S_\pi$ is the unitary representation of the symmetric group on $K$
fermion modes.  We determine the evolution for the system
interacting with the $j=1$ mode of the reference frame at each step,
and then apply the shuffling operator to the state of the reference
frame to obtain the state corresponding to an interaction with a
random mode as in Eqs.~(\ref{eq:fermionW1}-\ref{eq:fermionW3}).

We first apply a $\pi/2$-pulse by evolving with the Hamiltonian
$\hat{H}^{(j=1)}$ for time $t = \pi/(2\kappa)$ (step 1), then freely
evolving off-resonance for time $\tau$ (step 2), followed by another
$\pi/2$-pulse (step 3), yielding at each step $i=0,1,2,3$
\begin{equation}
  \Bigl[ \epsilon |A\rangle\langle A| \otimes |0\}\{0|
  +(1-\epsilon) |\Psi_i\rangle_L\langle\Psi_i| \Bigr]
  \otimes \sigma^{\otimes K-1} \,,
\end{equation}
where
\begin{align}
  \label{eq:fermionpsi0}
  |\Psi_0\rangle_L &= |A\rangle |1\} \,, \\
  \label{eq:fermionpsi1}
  |\Psi_1\rangle_L &= \frac{1}{\sqrt{2}} \Bigl( |A\rangle |1\}
  - i |M\rangle |0\} \Bigr) \,, \\
  \label{eq:fermionpsi2} |\Psi_2\rangle_L &= \frac{1}{\sqrt{2}} \Bigl( |A\rangle |1\}
  - i e^{-i\phi}|M\rangle |0\} \Bigr) \,, \\
  \label{eq:fermionpsi3} |\Psi_3\rangle_L &= \cos(\phi/2) |A\rangle |1\} - \sin(\phi/2) |M\rangle
  |0\}\,.
\end{align}
Because the mode with which the system interacts is unknown, the
state of the laboratory at any particular stage ``$i$'' of the
experiment may be written
\begin{equation}\label{eq:fermionpsisym}
  W_{i L} =
  \mathcal{S}\Bigl[ \bigl(\epsilon |A\rangle\langle A| \otimes |0\}\{0|
  +(1-\epsilon) |\Psi_i\rangle_L\langle\Psi_i|\bigr)
  \otimes \sigma^{\otimes K{-}1}\Bigr]\,,
\end{equation}
where, in a slight abuse of notation, it is understood that
$\mathcal{S}$ acts only on the $K$ reference-frame modes and as
identity on the system.  It is straightforward to show that these
states are equivalent to those presented in
Eqs.~(\ref{eq:fermionW1}-\ref{eq:fermionW3}).

The probabilities for measuring an atom or a molecule after the
final step are, respectively,
\begin{align}
  \label{eq:SingleFermionModeProbA}
  p_A &= {\rm Tr}\bigl[ W_{3L} |A\rangle\langle A| \bigr]
  = \epsilon + (1-\epsilon)\cos^2(\phi/2)\,, \\
  p_M &= {\rm Tr}\bigl[ W_{3L} |M\rangle\langle M| \bigr]
  = (1-\epsilon)\sin^2(\phi/2)\,,
  \label{eq:SingleFermionModeProbM}
\end{align}
yielding a visibility of $V=(1-\epsilon)$.  Thus, for small
$\epsilon$, we can achieve high visibility for the Ramsey fringes.

Along with achieving a high visibility in the limit $\epsilon\to 0$,
this result possesses another analogy with the bosonic experiment of
Sec.~\ref{sec:AtomMol}: in the limit $K \to \infty$, the state of
the reference frame is left undisturbed throughout the experiment.
Specifically, the reduced density operator of the reference frame
remains $\rho_0^{\rm rf}$ of Eq.~\eqref{eq:multimodefermioninitial}
after the experiment, regardless of what measurement outcome (atom
or molecule) is obtained.  We provide a proof of this fact in the
Appendix.

\subsubsection{Interpreting the results}

In the limit discussed above, the state of the reference frame
described above is undisturbed throughout the interference
experiment. Specifically, if the system is prepared in the state
$|A\rangle$ and the measurement result $|M\rangle$ is obtained, the
state of the reference frame is undisturbed even though a fermionic
atom has been removed from the reference frame.  This property is
shared with the single-mode bosonic atom state of
Eq.~\eqref{eq:coherentmixture2}.

It is illustrative to compare the single-mode bosonic atom state of
Eq.~\eqref{eq:coherentmixture2} and the multi-mode fermionic atom
state of Eq.~\eqref{eq:multimodefermioninitial}, and to address
possible concerns of a skeptic who questions whether the latter is a
good generalization of the former for the purposes of a Ramsey
experiment.

Bosons admit coherent states that are eigenstates of the
annihilation operator; such coherent states yield good visibility in
an interference experiment, and are also undisturbed by the
interactions.  Thus, one way to explain the fact that the state
\eqref{eq:coherentmixture2} is undisturbed by the interference
experiment is to note that every coherent state would be
undisturbed, so the convex sum of them will also be undisturbed. One
might naively expect that these properties \emph{cannot} be
generalized to fermionic states, because it is not possible to
define pure states of (single- or multi-mode) fermionic systems that
are eigenstates of an annihilation operator.

However, no convex decomposition should be preferred over others. We
can also view the mixed bosonic state of
Eq.~\eqref{eq:coherentmixture2} as a Poissonian distribution of
number states, as in Eq.~\eqref{eq:coherentmixture}. In that case,
it is the large uncertainty in total number that explains why the
mixed state is undisturbed.  One can appeal to this same sort of
explanation in the fermionic case.  A skeptic might still claim that
there is difference, namely, that `really' the bosonic system is in
a coherent state, and it is the non-disturbance to this `real' state
that is significant.  However, to make this statement is to commit
the partition ensemble fallacy~\cite{Kok00}.

Another possible distinction between these two states that a skeptic
might claim to be important is that, in the fermionic case, there is
classical information (the integer $j$) indicating the mode of the
reference frame with which the system interacted.  If one considers
the state of this mode, then it is highly disturbed by the
interaction.  However, the classical uncertainty about \emph{which}
mode was the mode of interaction ensures that the reference frame,
as a whole, is undisturbed.  This classical uncertainty ensures
that, if the experiment was repeated many times, there is only a
vanishing probability that the system will interact with the same
mode more than once; thus, visibility is maintained for many runs of
the experiment.

There are indeed significant differences between the bosonic and
fermionic states described here, in terms of how they can be used as
reference frames.  We discuss one such difference in
Sec.~\ref{subsec:Lifting}. However, we emphasize that, if one takes
an operational view of these Ramsey-type interference experiments,
then the bosonic and fermionic examples are completely equivalent in
that they produce high-visibility fringes for potentially many
repetitions of the experiments.

\subsection{A relational description}

This experiment exhibits high-visibility Ramsey fringes,
demonstrating coherence between a fermionic atom and a bosonic
molecule.  However, as with the experiment illustrated in
Sec.~\ref{subsec:QRF}, the reduced density operator for the system
at all times during the experiment is diagonal in the $|A\rangle$,
$|M\rangle$ basis.  To observe the coherence, one must instead look
to a relational description, which we now develop in analogy to that
presented in Sec.~\ref{sec:relational}.

First, we introduce some simplifying notation.  Let $\vec{n}$ be a
vector in $(\mathbb{Z}_2)^{\otimes K}$, i.e., a $K$-dimensional
vector where each element $n^{(i)}$ is either zero or one.  Then
$|\vec{n}\} = |n^{(1)},n^{(2)},\ldots,n^{(K)}\rangle$ is a Fock
state of $K$ fermionic modes.

Now, note that the states $W_{iL}$ of Eq.~\eqref{eq:fermionpsisym}
are block-diagonal in the eigenspaces of \emph{total} number of
atoms of type 2 (counting any atoms of type 2 which are bound into
molecules as well).  Defining a total type-2 atom number operator
\begin{equation}\label{eq:FermionTotalN2}
    \hat{N}_2 = \hat{n}_M + \sum_{j=1}^K \hat{n}_2^{(j)} \,,
\end{equation}
where $\hat{n}_M = \hat{b}_M^\dag \hat{b}_M$ and $\hat{n}_2^{(j)} =
\hat{f}_2^{(j)\dag}\hat{f}_2^{(j)}$, we can express
\begin{equation}\label{eq:WiLareblockdiagonal}
    W_{iL} = \sum_{N=0}^K c_N^{(K)}(1{-}\epsilon) W_{iL}^{(N)} \,,
\end{equation}
where $c_N^K(\cdot)$ is defined in Eq.~\eqref{eq:binomialdef} and where the states $W_{iL}^{(N)}$ are eigenstates of $\hat{N}_2$ with
eigenvalue $N$.  In the limit $K\to\infty$, $\epsilon \to 0$ with
$K\epsilon$ fixed, this expression becomes
\begin{equation}\label{eq:WiLareblockdiagonalPoisson}
    \lim W_{iL} = \sum_{N=0}^\infty p_N(\bar{n}) W_{iL}^{(N)} \,,
\end{equation}
with $\bar{n}=K(1-\epsilon)$, with the states $W_{iL}^{(N)}$ given
by
\begin{equation}
  \label{eq:FermionWiln}
  W_{iL}^{(N)} = \mathcal{S}\Bigl[ |\Psi_i\rangle_L\langle\Psi_i|
  \otimes |\vec{n}_{N-1}\} \{ \vec{n}_{N-1} |\Bigr]\,,
\end{equation}
where $\vec{n}_{N-1}$ is the vector $(1,1,\ldots,1,0,\ldots,0)$ with
$N-1$ ones and $K-N$ zeros, $\mathcal{S}$ is the shuffling operator
of Eq.~\eqref{eq:Shuffling}, and the states $|\Psi_i\rangle_L$ are
given by Eqs.~(\ref{eq:fermionpsi0})-(\ref{eq:fermionpsi3}).  The
fact that the states $|\Psi_i\rangle_L$ are independent of $N$
suggests an alternate partitioning of the Hilbert space, as follows.

First note that our current tensor product structure for the Hilbert
space, in terms of modes, is associated with the eigenvalues of the
number operators
\begin{equation}\label{eq:ModeNumberOperators}
    \hat{n}_M\,,\ \hat{n}_2^{(1)}\,, \ \hat{n}_2^{(2)}\,,\ldots,\
    \hat{n}_2^{(K)}\,.
\end{equation}
We choose the following different set of commuting operators:
\begin{equation}\label{eq:RelationalNumberOperators}
    \hat{n}_M\,,\quad \hat{N}_2\,,\quad \hat{r}_2^{(i)}
    \equiv \hat{n}_2^{(i)}-\hat{n}_2^{(i{-}1)}\,, \ i=2,\ldots,K\,.
\end{equation}
We note that the states $|A\rangle|\vec{n}\}$ and
$|M\rangle|\vec{n}\}$ are also eigenstates of this new set of
operators, specifically,
\begin{align}\label{eq:EigenstatesN2Fermion}
    \hat{n}_M|A\rangle|\vec{n}\} &= 0\,,\\
    \hat{n}_M|M\rangle|\vec{n}\}&= |M\rangle|\vec{n}\} \,,\\
    \hat{N}_2|A\rangle|\vec{n}\}
    &= \bigl(\sum_i n^{(i)} \bigr) |A\rangle|\vec{n}\} \,, \\
    \hat{N}_2|M\rangle|\vec{n}\}
    &= \bigl(1+\sum_i n^{(i)} \bigr) |M\rangle|\vec{n}\}\,,\\
    \hat{r}_2^{(i)}|A\rangle|\vec{n}\}
    &= (n^{(i)}-n^{(i{-}1)}) |A\rangle|\vec{n}\} \,, \\
    \hat{r}_2^{(i)}|M\rangle|\vec{n}\}
    &= (n^{(i)}-n^{(i{-}1)}) |M\rangle|\vec{n}\} \,.
\end{align}

We note that in the limit $\epsilon\to 0$, the states we consider
have no support on the vector $|A\rangle|0,0,\ldots,0\}$ and thus we
can focus our attention on the subspace $\mathcal{H}'_L$ of
$\mathcal{H}_L$ that is orthogonal to this vector.

We introduce a new tensor product structure on $\mathcal{H}'_L$ as
follows. We define a two-dimensional Hilbert space
$\mathcal{H}_{\text{rel}}$ with an orthonormal basis denoted by
$|A\rangle_{\text{rel}}$ and $|M\rangle_{\text{rel}}$, corresponding
respectively to $n_M=0$ (no molecules) and $n_M=1$ (one molecule).
We call this the \emph{relational} Hilbert space. We also define a
Hilbert space $\mathcal{H}_{\text{gl}}$ which has an orthonormal
basis labelled by $(N_2,\vec{r}_2)$, where $\vec{r}_2$ is the vector
consisting of the eigenvalues of the operators $\hat{r}_2^{(i)}$ for
$i=2,3,\ldots,K$.  These labels are defined for $N_2 \ge 1$, and
$\vec{r}_2$ consistent with this total type-2 atom number $N_2$.  We
call this the \emph{global} Hilbert space.  Then, because the
spectra of $\hat{n}_M$ and that of the operators $\hat{N}_2$ and
$\hat{r}_2^{(i)}$ are independent, we have a virtual tensor product
structure
\begin{align}
  \mathcal{H}'_L\simeq
  \mathcal{H}_{\text{rel}}\otimes\mathcal{H}_{\text{gl}}\,,
\end{align}
which is defined in terms of a linear map from $\mathcal{H}_S
\otimes \mathcal{H}_R$ to $\mathcal{H}_{\rm rel} \otimes
\mathcal{H}_{\rm gl}$ in terms of their respective basis states as
\begin{align}
  |A\rangle|\vec{n}\} &\mapsto |A\rangle_{\text{rel}}
  \bigl|N_2=\bigl(\sum_i n_2^{(i)}\bigr),\vec{r}_2\bigr\rangle_{\text{gl}}\,, \\
  |M\rangle|\vec{n}\} &\mapsto |M\rangle_{\text{rel}}
  \bigl|N_2=\bigl(1+\sum_i
  n_2^{(i)}\bigr),\vec{r}_2\bigr\rangle_{\text{gl}}\,,
\end{align}
where $r_2^{(i)} = n_2^{(i)} - n_2^{(i{-}1)}$.

The states $W_{iL}$ of Eq.~\eqref{eq:WiLareblockdiagonalPoisson} are
expressed in terms of this new tensor product structure as
\begin{equation}
  W_{iL} = |\Psi_i\rangle_{\rm rel}\langle \Psi_i| \otimes \rho_{\rm gl} \,,
\end{equation}
where the states on the relational Hilbert space are pure, given by
\begin{align}
  \label{eq:fermionpsi0rel}
  |\Psi_0\rangle_{\rm rel} &= |A\rangle_{\rm rel} \,, \\
  \label{eq:fermionpsi1rel}
  |\Psi_1\rangle_{\rm rel} &= \frac{1}{\sqrt{2}} \Bigl( |A\rangle_{\rm rel}
  - i |M\rangle_{\rm rel} \Bigr) \,, \\
  \label{eq:fermionpsi2rel} |\Psi_2\rangle_{\rm rel}
  &= \frac{1}{\sqrt{2}} \Bigl( |A\rangle_{\rm rel}
  - i e^{-i\phi}|M\rangle_{\rm rel} \Bigr) \,, \\
  \label{eq:fermionpsi3rel} |\Psi_3\rangle_{\rm rel}
  &= \cos(\phi/2) |A\rangle_{\rm rel} - \sin(\phi/2)
  |M\rangle_{\rm rel} \,.
\end{align}
The state $\rho_{\rm gl}$ on $\mathcal{H}_{\rm gl}$ is identified
via the following observations about the shuffling operation
$\mathcal{S}$. First, if a state $\rho$ has support entirely within
an eigenspace of the operators $\hat{N}_2$ and $\hat{n}_M$ with
eigenvalues $N$ and $n_m$, respectively, then $\mathcal{S}[\rho]$
also has support entirely within this same eigenspace; intuitively,
this is because symmetrization does not alter the total type-2 atom
number, or whether a type-2 atom is bound into a molecule or not.
(Recall that the shuffling operation acts only on the state of the
reference frame.)  Second, the shuffling operation completely
randomizes the eigenvalues of the operators $\hat{r}^{(i)}$, i.e.,
if $\rho$ on $\mathcal{H}'_L$ has support entirely within an
eigenspace of the operators $\hat{N}_2$ and $\hat{n}_M$ with
eigenvalue $N$ and $n_M$, respectively, then $\mathcal{S}[\rho]$ is
uniform mixture of \emph{all} states on $\mathcal{H}'_L$ that are
eigenstates $\hat{N}_2$ and $\hat{n}_M$ with the same eigenvalues.

Thus, the state $\rho_{\rm gl}$ is given by
\begin{equation}
  \rho_{\rm gl} = \sum_{N=0}^{\infty} p_{N}(\bar{n}) \Sigma_N\,,
\end{equation}
where $\Sigma_N$ is the completely mixed state on the eigenspace of
$\hat{N}_2$ in $\mathcal{H}_{\rm gl}$ with eigenvalue $N$.

Thus we see that having fermionic atoms does not present any new
difficulties compared to the case of bosonic atoms, so that one may
interpret the fermionic version of this interference experiment in
precisely the same way that we interpreted the bosonic version, as
described in Secs. \ref{subsec:Interpreting} and
\ref{sec:relational}.

\section{Discussion}

\subsection{Experimental Considerations}

The above descriptions of our proposed experiments are clearly
idealized and intended to illustrate the essential physics.  We now
address some of the issues that may arise in attempting to perform
our proposed experiments in real systems of ultracold bosonic or
fermionic atoms and molecules.

Both the bosonic and fermionic versions of our proposed experiment
require creation of mixtures of degenerate atoms of two different
species.  Such mixtures have now been created with a number of
different atomic species via the process of sympathetic
cooling~\cite{modugno2001,modugno2002,roati2002,hadzibabic2002,silber2005}.
In~\cite{modugno2001} a mixture of two BECs of different bosonic
species, $^{41}$K and $^{87}$Rb was created.  Furthermore the
location of Feshbach resonances in these two atomic species was
estimated in~\cite{simoni2003}.  We therefore consider these two
species as good candidates for implementing the bosonic version of
our experiment described in Sec.~\ref{sec:AtomMol}.  For the
fermionic version of the experiment described in
Sec.~\ref{sec:bosonfermioncoherence} we require a mixture of two
different fermionic species.  While mixtures of bosonic and
fermionic species are
common~\cite{modugno2002,roati2002,hadzibabic2002,silber2005}, to
our knowledge simultaneous degeneracy of two different fermionic
atom species has not yet been achieved experimentally.  However, a
degenerate mixture of two different spin states of the same
fermionic atom has been created~\cite{kohl2005}, which would suffice
as the two distinguishable fermionic species for our purposes.

Clearly one of the issues in performing the experiment, once
appropriate atomic species have been chosen, is detection.  We seem
to require the ability to perform a projective measurement of a
single atom or molecule in order to observe the interference
pattern, Eq.~\eqref{eq:perfectinterference}.  The issue of detecting
single atoms also arises in the context of quantum information
processing (QIP) with neutral atoms in optical lattices and magnetic
microtraps (see~\cite{treutlein2006} for a review of experimental
progress, and references therein).  It may be possible to perform
the projective measurement of atom or molecule using the fiber-based
Fabry-Perot resonators described in~\cite{treutlein2006}.

Alternatively, one may attempt to perform many copies of the
experiment simultaneously by beginning with an optical lattice
containing atoms of the first species in the Mott-insulator regime
at unit filling, so that atoms in different lattice sites are
essentially non-interacting.  Again, the creation of optical
lattices containing precisely one atom per site has been considered
in the context of QIP with neutral atoms~\cite{treutlein2006}.  We
would then seem to require a separate BEC of the second atomic
species at each site to create distinct copies of the same
experiment.  However, this could be challenging experimentally and given that we are working in the classical limit it may suffice to use a single BEC with a spatial profile that overlaps the entire lattice\footnote{The variation in phase over the profile of the BEC is
unlikely to be relevant because, as we discussed above, the absolute
phase of the BEC is irrelevant to the interference pattern.}. The interactions could be implemented by tuning an external
magnetic field (uniform over the lattice) onto the Feshbach
resonance.   Free evolution could be implemented by switching off
the external magnetic field for the desired period of time.  At the
end of the experiment standard techniques should suffice to detect
the number of atoms of the first species present in the lattice
(molecules being typically much more difficult to detect), and we
would expect to see an interference pattern in this number as a
function of the free-evolution time.

Next, from a theoretical perspective, one may question whether the
single-mode description used in Sec.~\ref{sec:AtomMol} is adequate
for the BEC.  For small BECs there is some ``quantum depletion''
whereby some atoms do not occupy the condensate, even at zero
temperature, due to interactions between the
atoms~\cite{dalfovo1999}.  However this effect is small (a few
percent or less) in most current experiments, and in any case we are
concerned with the limit of a large BEC --- precisely the limit in
which the mean-field dynamics given by the Gross-Pitaevski equation
become exact.

For the experiment involving fermions described in
Sec.~\ref{sec:bosonfermioncoherence} both preparation of the initial
state and implementation of the interaction are likely to be far
more difficult.  We do not have any concrete suggestions for how this
experiment might be performed, however we note that the initial
state, Eq.~\eqref{eq:multimodefermioninitial}, may be
well-approximated by fermions near the Fermi level at a small but
non-zero temperature, i.e.\ the different fermion modes are the
momentum modes close to the Fermi momentum.  Alternatively, it may
be more convenient to use spatially isolated modes, such as the
lattice sites in an optical lattice, as the multimode fermionic
reference state.  Fermionic atoms were confined to a three
dimensional optical lattice in~\cite{kohl2005}, so the reference
state~\eqref{eq:multimodefermioninitial} could perhaps be created by
trapping slightly fewer atoms than the number of lattice sites, so
that the vacant sites are randomly distributed.  One must also
address the issue of implementing the non-unitary coupling described
by Eqs.~(\ref{eq:fermionW1}-\ref{eq:fermionW3}).  It may be possible
to implement such a coupling using the Hamiltonian of
Eq.~\eqref{eq:CoherentFermionFeshbach} if the reference frame is
kept in thermodynamic equilibrium with a particle heat bath that can
replenish an exhausted fermionic mode.

Finally, another question that must be addressed concerns the
timescales and mechanisms for decoherence.  One might expect
interactions with background atoms and molecules and with
uncondensed thermal atoms to cause the superpositions to decohere,
but whether or not this would occur on a timescale shorter than is
necessary to observe the interference fringes has not been studied,
to our knowledge.

Thus there are clearly many experimental challenges to be overcome.
However many of the required elements have been demonstrated
individually. With the rapid pace of progress in the field of
ultracold atomic physics, where molecular condensates, degenerate
Fermi gases and Bose-Fermi mixtures are all topics of much current
interest, it is plausible that experiments of this type may be
performed in the near future.  This would open up the possibility of
experimentally investigating the role of superselection rules in
these systems.

\subsection{Lifting superselection rules}
\label{subsec:Lifting}

If a superselection rule is completely lifted by the existence of an
appropriate reference frame, meaning that one can in principle
perform any quantum operation as if the superselection rule did not
exist, then one would expect to be able to perform an experiment on
\emph{two} systems that exhibits a relative phase between these
systems, independent of the reference frame. Specifically, one could
generalize the methods we presented here to define a two-system
relational Hilbert space for the composite; states on this
relational Hilbert space would describe the two systems relative to
the single reference frame. One would expect that some degrees of
freedom in this relational Hilbert space describe the relation
between the two systems themselves, independent of the reference
frame. For example, one could perform measurements of observables
defined on the relational Hilbert space that provide information
about the relative phase of the two systems.

Consider the following experiment.  Two two-level systems such as
those described in this paper are initially prepared as single
atoms.  A $\pi/2$-pulse is performed on the first system followed by
a $\pi/2$-pulse on the second system (both pulses being
implemented by interaction with the reference frame), and then a
phase shift $\phi$ is applied to the second system.  One would
expect, then, that a measurement of the relative phase on the
relational Hilbert space could yield information about $\phi$.  (One
such measurement would be the two-outcome projection onto the
symmetric and antisymmetric subspaces of the relational Hilbert
space.  If the two systems are in phase, the symmetric outcome is
obtained with certainty.)  For the bosonic reference frames explored
in Sec.~\ref{sec:AtomMol}, it is straightforward to show that this
is indeed the case. However, for the fermion reference frame
introduced in Sec.~\ref{sec:bosonfermioncoherence}, such an
experiment would be completely insensitive to $\phi$. The reason for
this insensitivity is that, for the interactions describing the
$\pi/2$ pulses, each of the two systems would interact with a
\emph{different} random mode of the reference frame. As the
individual reference-frame modes are uncorrelated in phase, the two
systems would also be uncorrelated.

Thus, although our second example of a Ramsey interference
experiment exhibits a coherent superposition of a single fermionic
atom and a bosonic molecule, we do not currently foresee how such an
experiment can be directly generalized to create arbitrary
superposition states of multiple systems.  The form of reference
frame used in Sec.~\ref{sec:bosonfermioncoherence} does not lift the
superselection rule, but provides only a demonstration of a
violation of this rule in a single system.

\section{Conclusions}

We conclude by responding to some anticipated objections by a
skeptic who questions that our gedanken-experiments would exhibit a
coherent superposition of a single atom and a molecule.  Suppose a
skeptic asserted that the only adequate description of the
atom-molecule interference experiments is the one wherein the BEC is
internalized, that is, treated quantum mechanically. She could
appeal to the fact that there is no coherence between the states
$|A\rangle$ and $|M\rangle$ on $\mathcal{H}_S$ in this description
to argue that a coherent superposition of atom and molecule had
actually \emph{not} been generated. We would of course disagree with
the assessment that internalizing the reference frame is the only
way to obtain an adequate description of the experiment, but leaving
this aside, one can respond to such a skeptic by noting that this
argument would also apply to the Ramsey experiment outlined in
Sec.~\ref{sec:Ramsey}. The reason is that when the EM field in the
Ramsey experiment is treated quantum mechanically, then the reduced
density operator on the Hilbert space of the atom has no coherence
between the internal states $|A\rangle$ and $|B\rangle$.

Thus, a skeptic could deny that we have demonstrated the possibility
of having a coherent superposition of atom and molecule, but then
she would also have to deny that the Ramsey interference experiment
demonstrates coherence between two internal states of an atom. In
fact, the sort of argument the skeptic is presenting can be applied
to essentially \emph{any} degree of freedom that requires a
reference frame for its definition.

For instance, suppose one internalized the reference frame for
spatial orientation or for the spatial position of a system.  Then
by exactly analogous arguments to those presented here, the reduced
density operator on the Hilbert space of the system would be found
to be an \emph{incoherent} sum of angular momentum or linear
momentum eigenstates. Thus, a dogmatic insistence on the necessity
of internalizing reference frames would lead one in this context to
conclude that it is impossible to prepare coherent superpositions of
angular momentum or linear momentum eigenstates. But although the
latter quantities are conserved, no one feels that it is fruitful to
insist on a superselection rule for them\footnote{An exception is
found in some approaches to quantum gravity.  The argument for why
such superselection rules should be in force for descriptions of the
quantum state of the universe is precisely because there are no
external references frames in such a description.}.

The only real difference we can see between reference frames for
spatial orientation, position, or the phase of the internal state of
an atom, on the one hand, and for the phase conjugate to charge,
atom number, or univalence on the other, is that reference frames of
the first sort are ubiquitous while those of the latter sort are
difficult to prepare.  Any rigid object can act as a reference frame
for spatial position, whereas a reference for the phase conjugate to
atom number presumably requires one to have succeeded in the
experimentally challenging task of achieving and maintaining
Bose-Einstein condensation for that atomic species.

Along these lines, we note that the recent demonstrations of Ramsey
fringes in so-called ``superconducting qubits'' in a
single-Cooper-pair box configuration~\cite{Nak99} is analogous in
many ways to the experiments we propose here (see \cite{beenakker2005} for related theory).  This experiment demonstrates coherence between states of differing charge, i.e., of
states of a superconducting island which differ by charge $2e$ (the
charge of a single Cooper pair). The reference frame for this
system, which is used to implement the Ramsey pulses, is a nearby
superconductor, the state of which -- a BCS ground state -- has an
accurate and successful description as a classical field.  This
superconducting qubit experiment can be interpreted as violating a
superselection rule for charge~\cite{aharonov1967,Ker74}, in direct
analogy with the way that the interference experiments proposed here
can be interpreted as violating a superselection rule for
atom-number.

\begin{acknowledgments}
The authors gratefully acknowledge Howard Wiseman for pointing out
that the univalence superselection rules has not been lifted,
Dominic Berry for suggesting the method used to prove
Eq.~\eqref{eq:FidLimit}, and Andrew Doherty and David Poulin for
helpful discussions.   SDB acknowledges support from the Australian
Research Council.  TR acknowledges support from the Engineering and
Physical Sciences Research Council of the United Kingdom.  RWS
acknowledges support from the Royal Society.
\end{acknowledgments}

\appendix*

\section{Proof that the state of the fermionic reference frame is
undisturbed}

The state of the laboratory after the second $\pi/2$-pulse is given
by Eq.~\eqref{eq:fermionpsisym} using Eq.~\eqref{eq:fermionpsi3} as
\begin{equation}
  W_{3L} =
  \mathcal{S}\Bigl[ \bigl(\epsilon |A\rangle\langle A| \otimes |0\}\{0|
  +(1-\epsilon) |\Psi_3\rangle_L\langle\Psi_3|\bigr) \otimes \sigma^{\otimes K-1}
  \Bigr]\,,
\end{equation}
with
\begin{equation}
  |\Psi_3\rangle_L = \cos(\phi/2) |A\rangle |1\} - \sin(\phi/2) |M\rangle
  |0\}\,.
\end{equation}
Consider the postselected states of the reference frame conditioned
on measuring an atom or a molecule, given by
\begin{equation}\label{eq:PostSelectedStates}
    \rho^{\rm rf}_A = \frac{\langle A|W_{3L}|A\rangle}{p_A}\,, \quad
    \rho^{\rm rf}_M = \frac{\langle M|W_{3L}|M\rangle}{p_M}\,,
\end{equation}
which occur with probabilities given by
Eqs.~\eqref{eq:SingleFermionModeProbM} and
\eqref{eq:SingleFermionModeProbA}.  These postselected states are
diagonal in the fermion-number basis and can be expressed as
\begin{align}
  \nonumber \rho^{\rm rf}_A =& \frac{1}{p_A} \sum_{n{=}0}^{K{-}1}
  c^{K{-}1}_n(1{-}\epsilon) \Bigl[
  \epsilon \mathcal{S}\bigl[|1\}^{\otimes n} |0\}^{K{-}n} \bigr] \nonumber \\
  &\ +(1-\epsilon) \cos^2(\phi/2) \mathcal{S}\bigl[|1\}^{\otimes n+1} |0\}^{K{-}n{-}1} \bigr] \Bigr] \,, \\
  \rho^{\rm rf}_M =& \sum_{n{=}0}^{K{-}1} c^{K{-}1}_n(1{-}\epsilon)
  \mathcal{S}\bigl[|1\}^{\otimes n} |0\}^{K{-}n} \bigr]\,.
\end{align}
The postselected state conditioned on an atom, $\rho^{\rm rf}_A$,
can be written as a sum of two terms,
\begin{equation}
  \label{eq:rhoAtwoterms} \rho^{\rm rf}_A = \frac{1}{p_A}\left( \epsilon \rho^{\rm rf}_M
  + (1-\epsilon) \cos^2(\phi/2)  \tilde{\rho}^{\rm rf}_A \right)\,,
\end{equation}
where we define
\begin{equation}
  \tilde{\rho}^{\rm rf}_A = \sum_{n{=}1}^{K} c^{K{-}1}_{n{-}1}(1{-}\epsilon)
  \mathcal{S}\bigl[|1\}^{\otimes n} |0\}^{K{-}n} \bigr]\,.
\end{equation}

We now show that the postselected states $ \rho^{\rm rf}_{A}$ and
$\rho^{\rm rf}_{M}$ are indistinguishable from each other, and from
the initial state $\rho^{\rm rf}_{0}$, in the limit of $K\to
\infty$ for a fixed $\epsilon > 0$.  The intuition for expecting
this result is that the total number of fermions in the reference
frame is indeterminate (a binomial distribution) so
that in the limit of large number of fermions, distributed in an
even larger number of modes, we will not be able to tell if one of
the modes has interacted with the system and possibly lost one
fermion.  The random coupling of the map $\mathcal{E}_U$ ensures
that we do not know which fermion mode to look at.

From~\eqref{eq:rhoAtwoterms} it is clear that the non-trivial step
is to show that $\tilde{\rho}^{\rm rf}_A$ and $\rho^{\rm rf}_M$ are
indistinguishable in the limit.  We will use the fidelity, defined
as~\cite{nielsen2000a}
\begin{equation}
F(\rho,\sigma) = \Tr \left(\sqrt{\rho^{1/2} \sigma \rho^{1/2}}\right)
\end{equation}
as our measure of the indistinguishability of two states $\rho$ and
$\sigma$, where $F=1$ implies that the states are completely
indistinguishable.  For states $\rho = \sum_i r_i |i\rangle\langle
i|$ and $\sigma = \sum_i s_i |i\rangle\langle i|$ diagonal in the
same basis the fidelity reduces to
\begin{equation}
F(\rho,\sigma) = \sum_i \sqrt{r_i s_i}.
\end{equation}
Therefore the fidelity of these two states is
\begin{align}
  F(&\tilde{\rho}^{\rm rf}_A,\rho^{\rm rf}_M) = \sum_{n{=}1}^{K{-}1} \sqrt{c^{K{-}1}_{n{-}1} c^{K{-}1}_n}\,, \\
  &= 1-\frac{c^{K{-}1}_0}{2}-\frac{c^{K{-}1}_{K{-}1}}{2} 
  -\frac{1}{2}\sum_{n=1}^{K-1}\left(\sqrt{c^{K{-}1}_{n-1}}-\sqrt{c^{K{-}1}_n}\right)^2 \,, \\
  &\geq 1-\frac{c^{K{-}1}_0}{2}-\frac{c^{K{-}1}_{K{-}1}}{2} 
  -\frac{1}{2}\sum_{n=1}^{K-1} \left|\sqrt{c^{K{-}1}_{n-1}}-\sqrt{c^{K{-}1}_n} \right| \,, \\
  &= 1-c^{K{-}1}_0-c^{K{-}1}_{K{-}1}-c^{K{-}1}_{\rm max}\,, \\
  &\to 1 \,, \label{eq:FidLimit}
\end{align}
where $c^{K{-}1}_{\rm max} = \max_n c^{K{-}1}_n$ is the probability
of the most likely number of fermions (the floor or ceiling of the
mean) and we have not written explicitly the dependence on
$(1-\epsilon)$ of the binomial probabilities.

Therefore from~\eqref{eq:rhoAtwoterms} it is clear that the fidelity
$F(\tilde{\rho}^{\rm rf}_A,\rho^{\rm rf}_M)$ between the two
postselected states also approaches $1$ in the limit $K \to \infty$.
Similar calculations show the same result for the postselected
states of the reference frame at either of the intermediate times in
the experiment, Eqs.~\eqref{eq:fermionpsi1} or
\eqref{eq:fermionpsi2} substituted into
Eq.~\eqref{eq:fermionpsisym}.  Furthermore either of the
postselected states of the reference frame at any time during the
experiment can be shown to have fidelity approaching $1$ with the
initial state $\rho^{\rm rf}_0$ in the same limit.  Therefore we
have the result we claimed --- the reference frame is undisturbed by
the interaction in the limit of a large number of modes.

\end{document}